\documentclass[trackchanges,twocolumn]{aastex701}

\usepackage{enumitem, amsmath, xspace, placeins, multirow, graphicx}
\usepackage[dvipsnames]{xcolor}
\setlist[itemize]{itemsep=6pt, parsep=0pt, leftmargin=\parindent}

\newcommand{\smolbullet}{$\boldsymbol{\cdot}$}
\newcommand{\halpha}{\ensuremath{\rm{H}\alpha}\xspace}

\newcommand{\paalpha}{\ensuremath{\rm{Pa}\alpha}\xspace}
\newcommand{\ci}{[C{\scriptsize I}]\xspace}
\newcommand{\cii}{[C{\scriptsize II}]\xspace}
\newcommand{\oiii}{[O{\scriptsize III}]\xspace}
\newcommand{\jwst}{{\it JWST}\xspace}
\newcommand{\hst}{{\it HST}\xspace}

\defcitealias{marrone_galaxy_2018}{M18}
\defcitealias{arribas_ga-nifs_2024}{A24}
\defcitealias{alvarez-marquez_mirijwst_2023}{AM23}
\defcitealias{spilker_chaotic_2022}{S22}

\shorttitle{Diverse Stellar Components of SPT0311$-$58}
\shortauthors{Wirag et al.}

\begin{document}

\title{Diverse Stellar Components Resolved by \jwst in the Massive Galaxy SPT0311$-$58}


\author[orcid=0009-0000-9036-4323, sname=Wirag]{Briana Wirag}
\affiliation{Department of Physics and Astronomy and George P. and Cynthia Woods Mitchell Institute for Fundamental Physics and Astronomy, Texas A\&M University, 4242 TAMU, College Station, TX 77843-4242, USA}
\email[show]{wirag@tamu.edu}

\author[orcid=0000-0002-3272-7568, sname=Birkin]{Jack E. Birkin}
\affiliation{Department of Physics and Astronomy and George P. and Cynthia Woods Mitchell Institute for Fundamental Physics and Astronomy,
Texas A\&M University, 4242 TAMU, College Station, TX 77843-4242, USA}
\email{jbirkin@tamu.edu}

\author[orcid=0000-0003-3256-5615, sname=Spilker]{Justin S. Spilker}
\affiliation{Department of Physics and Astronomy and George P. and Cynthia Woods Mitchell Institute for Fundamental Physics and Astronomy,
Texas A\&M University, 4242 TAMU, College Station, TX 77843-4242, USA}
\email{jspilker@tamu.edu}

\author[orcid=0000-0002-6922-469X, sname=Zhou]{Dazhi Zhou}
\affiliation{Department of Physics and Astronomy, University of British Columbia, 6224 Agricultural Road, Vancouver, BC, Canada}
\email{dzhou@astro.ubc.ca}

\author[orcid=0000-0002-6290-3198, sname=Aravena]{Manuel Aravena}
\affiliation{Instituto de Estudios Astrof\'{\i}cos, Facultad de Ingenier\'{\i}a y Ciencias, Universidad Diego Portales, Av. Ej\'ercito 441, Santiago, Chile}
\affiliation{Millenium Nucleus for Galaxies (MINGAL)}
\email{manuel.aravenaa@mail.udp.cl}

\author[orcid=0000-0002-0517-9842, sname=Archipley]{Melanie Archipley}
\affiliation{Department of Astronomy and Astrophysics, University of Chicago, 5640 South Ellis Avenue, Chicago, IL, 60637, USA}
\affiliation{Kavli Institute for Cosmological Physics, University of Chicago, 5640 South Ellis Avenue, Chicago, IL, 60637, USA}
\email{archipleym@uchicago.edu}

\author[orcid=0000-0001-8706-1268, sname=Burgoyne]{James R. Burgoyne}
\affiliation{Department of Physics and Astronomy, University of Victoria, Victoria, Canada}
\email{jburgoyne@uvic.ca}

\author[orcid=0000-0002-4657-7679, sname=Cathey]{Jared Cathey}
\affiliation{Department of Astronomy, University of Florida, 211 Bryant Space Sciences Center, Gainesville, FL 32611, USA}
\email{jaredcathey@ufl.edu}

\author[orcid=0000-0002-8487-3153, sname=Chapman]{Scott C. Chapman}
\affiliation{Department of Physics and Atmospheric Science, Dalhousie University, Halifax, Halifax, NS B3H 3J5, Canada}
\email{scott.chapman@dal.ca}

\author[orcid=0000-0002-0933-8601, sname=Gonzalez]{Anthony H. Gonzalez}
\affiliation{Department of Astronomy, University of Florida, 211 Bryant Space Sciences Center, Gainesville, FL 32611, USA}
\email{anthonyhg@astro.ufl.edu}

\author[orcid=0009-0008-8718-0644, sname=Hill]{Ryley Hill}
\affiliation{Department of Physics and Astronomy, University of British Columbia, 6224 Agricultural Road, Vancouver, BC, Canada}
\email{ryleyhill@phas.ubc.ca}

\author[orcid=0009-0004-7059-8658, sname=LeVar]{Nicholas LeVar}
\affiliation{Department of Physics and Astronomy and George P. and Cynthia Woods Mitchell Institute for Fundamental Physics and Astronomy, Texas A\&M University, 4242 TAMU, College Station, TX 77843-4242, USA}
\email{nlevar@tamu.edu}

\author[orcid=0000-0001-7946-557X, sname=Phadke]{Kedar A. Phadke}
\affiliation{Department of Astronomy, University of Illinois, 1002 West Green Street, Urbana, IL, 61801, USA}
\affiliation{Center for AstroPhysical Surveys, National Center for Supercomputing Applications, 1205 West Clark Street, Urbana, IL, 61820, USA}
\affiliation{NSF-Simons AI Institute for the Sky (SkAI), 172 E. Chestnut St., Chicago, IL 60611, USA}
\email{kphadke2@illinois.edu}

\author[orcid=0009-0003-6250-1396, sname=Pillai]{Vismaya Pillai}
\affiliation{Department of Physics and Astronomy, University of British Columbia, 6224 Agricultural Road, Vancouver, BC, Canada}
\email{vpillai@student.ubc.ca}

\author[orcid=0000-0001-8598-064X, sname=Posses]{Ana C. Posses}
\affiliation{Department of Physics and Astronomy and George P. and Cynthia Woods Mitchell Institute for Fundamental Physics and Astronomy, Texas A\&M University, 4242 TAMU, College Station, TX 77843-4242, USA}
\email{anaposses@tamu.edu}

\author[orcid=0000-0002-3187-1648, sname=Sulzenauer]{Nikolaus Sulzenauer}
\affiliation{Max-Planck-Institut f\"{u}r Radioastronomie, Auf dem H\"{u}gel 69, Bonn, D-53121, Germany}
\email{nsulzenauer@mpifr-bonn.mpg.de}

\author[orcid=0000-0001-7192-3871, sname=Vieira]{Joaquin D. Vieira}
\affiliation{Department of Astronomy, University of Illinois, 1002 West Green Street, Urbana, IL, 61801, USA}
\affiliation{Center for AstroPhysical Surveys, National Center for Supercomputing Applications, 1205 West Clark Street, Urbana, IL, 61820, USA}
\affiliation{Department of Physics, University of Illinois, 1110 West Green Street, Urbana, IL, 61801, USA}
\email{jvieira@illinois.edu}

\author[orcid=0000-0001-7610-5544, sname=Vizgan]{David Vizgan}
\affiliation{Department of Astronomy, University of Illinois, 1002 West Green Street, Urbana, IL, 61801, USA}
\email{dvizgan2@illinois.edu}

\author[orcid=0000-0003-4678-3939, sname=Weiß]{Axel Weiß}
\affiliation{Max-Planck-Institut f\"{u}r Radioastronomie, Auf dem H\"{u}gel 69, Bonn, D-53121, Germany}
\email{aweiss@mpifr-bonn.mpg.de}

\author[orcid=0000-0002-4861-0081, sname=Zhang]{Nan Zhang}
\affiliation{Department of Physics, University of Illinois, 1110 West Green Street, Urbana, IL, 61801, USA}
\email{nanz6@illinois.edu}


\begin{abstract}

SPT0311$-$58 is comprised of two clumpy dusty star-forming galaxies inhabiting a $z=6.9$ protocluster core. Within a $16$\,kpc diameter, there are 26 spectroscopically confirmed members. The 12 highly dust-obscured clumps were identified by their rest-frame \cii emission, and the 14 less obscured clumps in the rest-frame ultraviolet (UV). We present new NIRCam and MIRI imaging from \jwst of the central regions, which respectively trace the rest-frame UV and near-infrared. Combining this imaging with supplementary data from the \emph{Hubble Space Telescope} and Atacama Large Millimeter Array (ALMA) provides the most complete picture of SPT0311$-$58 to date. Using 14 photometric bands and ALMA dust continuum to model the spectral energy distribution of SPT0311$-$58, we present well-informed stellar masses and star-formation rates (SFR) on local ($\sim1.5$\,kpc) and global scales. We find that SPT0311$-$58 has a total stellar mass of $M_\star=3.2^{+3.2}_{-1.6}\times10^{11}$\,M$_\odot$ and SFR of $1.5^{+0.89}_{-0.53}\times10^3$\,M$_\odot\,\text{yr}^{-1}$. We compare these measurements to simulations and known, rare classes of high-redshift objects observed by \jwst, namely massive $z\approx6$--$8.5$ Lyman-break galaxies, $z>6$ quasars, and other merging/clumpy $z\sim6$--$7$ systems. SPT0311$-$58 is comparable to $z>6$ quasars in both stellar mass and SFR, but is more extreme than the other reference populations for both quantities. Additionally, we compare our measurements with the star-formation histories (SFH) of five massive quiescent galaxies from the Early eXtragalactic Continuum and Emission Line Science (EXCELS) survey. We find that SPT0311$-$58 has a higher SFR than expected from the SFHs of the EXCELS galaxies at $z=6.9$, but is otherwise a plausible progenitor to massive quiescent galaxies.

\end{abstract}



\section{Introduction}\label{sec:intro}

Dusty star-forming galaxies (DSFGs) are among the most extreme systems in the Universe \citep{smail97, blain02, casey_dusty_2014}. These galaxies can have star-formation rates (SFR) reaching up to thousands of solar masses per year \citep{magnelli_2012, miettinen17, dudzeviciute20}, and they contribute significantly to the total SFR density out to $z$\,$\sim$\,4--5 \citep{zavala_evolution_2021}. DSFGs are highly luminous in the submillimeter regime because dust obscures and reprocesses rest-frame optical and ultraviolet (UV) stellar radiation into far-infrared/submillimeter emission. Previous studies have pushed the exploration of the dusty SFR density to higher redshifts by leveraging the negative $K$-correction associated with submillimeter wavelengths, since flux density in this regime remains constant with redshift as opposed to decreasing \citep{reuter_complete_2020, zavala_evolution_2021}.

DSFGs can exhibit incredibly high SFRs, but vary in their mass-weighted SFRs, or specific SFRs (sSFR). When compared to the star-forming main sequence (SFMS), some DSFGs are consistent with the high mass expectation of a typical star-forming galaxy \citep{michalowski_stellar_2012, da_cunha_alma_2015, elbaz18, dudzeviciute20}, while others are clear starbursts \citep{daddi_multiwavelength_2007, ma_stellar_2015}. These intense episodes of star formation play a crucial role in the buildup of stellar mass, and such episodes are also thought to occur in the early formation histories of massive quiescent galaxies \citep{toft_deep_2012, toft_submillimeter_2014}. Therefore, DSFGs have long been speculated to be viable progenitors of these massive elliptical/quiescent galaxies \citep{blain02, coppin08, simpson14, toft_submillimeter_2014, stach19, dudzeviciute20, birkin21}. The discovery of quiescent galaxies beyond $z>$\,3--4 \citep{glazebrook_massive_2017,forrest_extremely_2020, carnall_massive_2023, carnall_jwst_2024, glazebrook_massive_2024}, however, implies rapid formation between $z\sim6$\,--10 \citep{araya-araya_connection_2025}.

The frontier of detecting and characterizing DSFGs at such high redshifts involves exploiting gravitational lenses, which boost sensitivity and angular resolution, providing the means for much more detailed studies. While the brightest DSFGs are rare \citep{negrello_2007}, wide-field surveys conducted by e.g. the South Pole Telescope \citep[SPT;][]{vieira10, vieira_dusty_2013} and {\it Herschel}-SPIRE \citep{negrello10, wardlow13} have revealed hundreds of such systems. One of the most distant and remarkable DSFGs, and the focus of this paper, is SPT0311$-$58 \citep{strandet_ism_2017} at $z=6.9$. This lensed system was the highest redshift source identified in the SPT-SZ survey, and is still the most distant spectroscopically confirmed DSFG to date \citep{everett_millimeter-wave_2020, reuter_complete_2020}. Using the gas mass of SPT0311$-$58 to estimate the dark matter (DM) halo mass, \citet{marrone_galaxy_2018}, hereafter \citetalias{marrone_galaxy_2018}, predicted a halo mass on the order of $10^{12}$--$10^{13}$\,M$_\odot$, concluding that SPT0311$-$58 occupies one of the most massive DM haloes within the Epoch of Reionization (EoR) known to date.

Given its extraordinary nature, SPT0311$-$58 has been the subject of an extensive follow-up campaign with the Atacama Large Millimeter Array (ALMA). \citetalias{marrone_galaxy_2018} presented emission from the 160\,$\mu$m dust continuum, which resolved SPT0311$-$58 into two galaxies: SPT0311$-$58E and SPT0311$-$58W. The magnification for SPT0311$-$58E is $\mu_E=1.3$ and it is weakly distorted by the lensing galaxy. SPT0311$-$58W falls along a lensing caustic and is therefore strongly lensed with a moderate magnification of $\mu_W=2.2$. \citetalias{marrone_galaxy_2018} combined their ALMA measurements with optical-to-infrared (IR) measurements to constrain the spectral energy distribution (SED) of both sources, inferring a magnification-corrected SFR for the dusty component (SPT0311$-$58W) of $2900\pm1800$\,M$_\odot$\,yr$^{-1}$. This followed CO, \ci, and \cii observations by \cite{strandet_ism_2017}, confirming the spectroscopic redshift and implying an intrinsic molecular gas mass of $M_{\rm gas}=(3.3\pm1.9)\times10^{11}$\,M$_\odot$. Additionally, \citetalias{marrone_galaxy_2018} presented UV/optical photometry that revealed extended UV emission associated with SPT0311$-$58E. At the time, imaging UV emission directly was somewhat rare among studies of DSFGs given the high dust obscuration of these systems. The resolution, however, was insufficient to enable a more detailed characterization of the stellar component.

\begin{figure*}[t!]
    \centering
    \includegraphics[width=\linewidth]{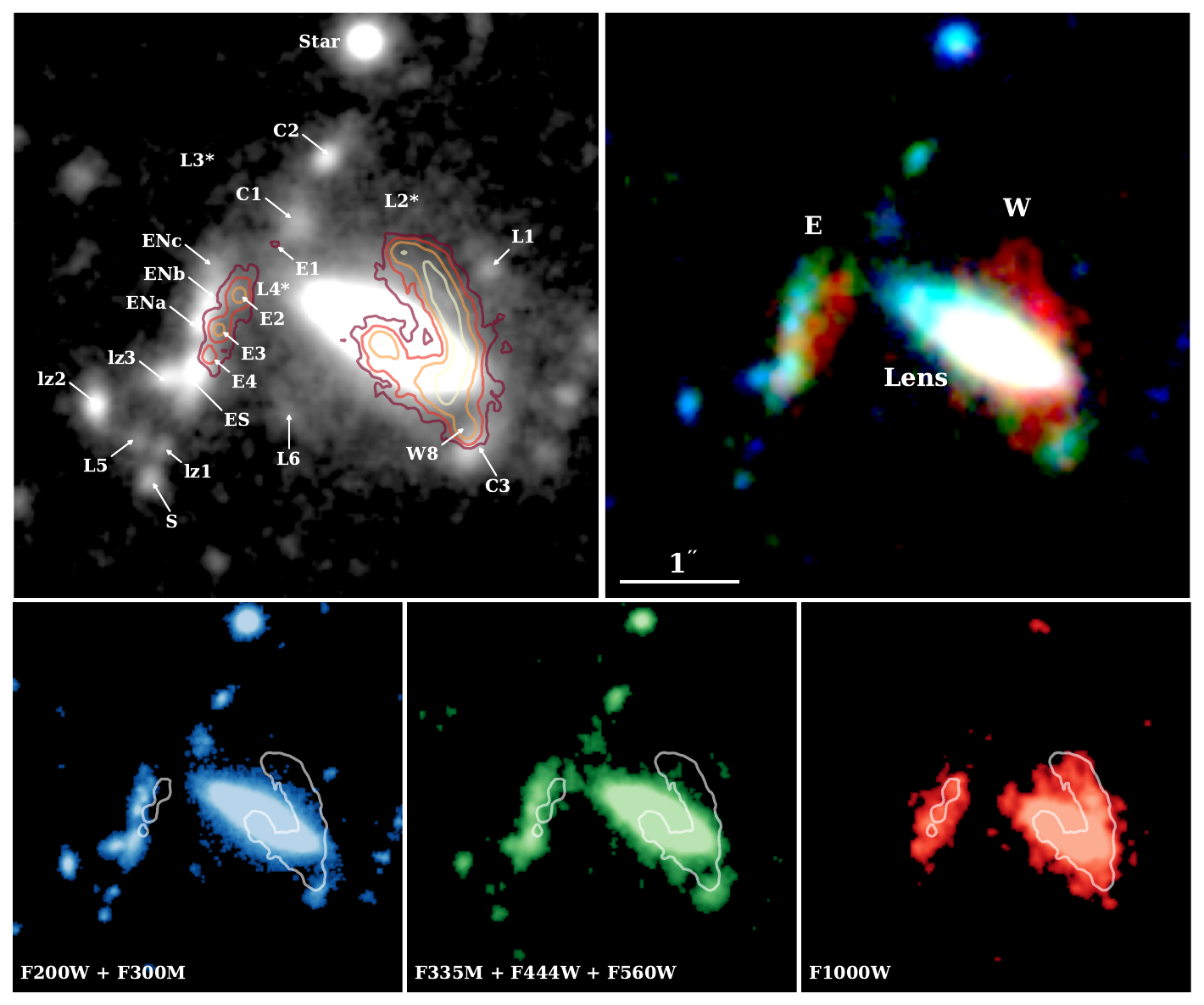}
    \caption{
    Top Left: Stacked NIRCam detection image (F200W+F300M+F335M+F444W) with ALMA 160$\mu$m contours denoting 6, 10, 20, and 50$\sigma$ overlaid. Known $z=6.90$ clumps identified by \citetalias{spilker_chaotic_2022} and \citetalias{arribas_ga-nifs_2024} are labeled. Our imaging does not detect the spectroscopically-identified sources L2, L3, or L4, which are marked with asterisks. Low-redshift interlopers lz1, lz2, and lz3 are marked, but we do not consider them further \citepalias{arribas_ga-nifs_2024}. Clumps W1 -- W7 \citepalias{spilker_chaotic_2022} are not labeled for clarity.
    Top Right: RGB image of SPT0311$-$58 using the filter combinations shown in the bottom row: NIRCam F200W $+$ F300M (blue), NIRCam F335M $+$ F444W $+$ MIRI F560W (green), and MIRI F1000W (red). The 10$\sigma$ 160$\mu$m contour is also overlaid in each of the bottom panels. The foreground lensing galaxy ($z=1.0343$, \citetalias{arribas_ga-nifs_2024}) has an average magnification of $\mu\sim2$. Most of SPT0311$-$58W is only visible in the MIRI images, and a spatial offset in SPT0311$-$58E between the NIRCam and MIRI images is apparent, likely due to differential dust extinction in this source.
    }
    \label{fig:rgb}
\end{figure*}

Thankfully, \jwst can probe the optical early Universe in unprecedented detail. Thus, early programs with \jwst have targeted DSFGs in attempts to resolve their UV and optical components \citep{colina_uncovering_2023, gillman24, polletta24, sun_jades_2024, hodge_aless-jwst_2025, rigby_templates_25}. With this work, we present new NIRCam and MIRI imaging data to resolve the stellar component of SPT0311$-$58 directly. We supplement this with data from the \emph{Hubble Space Telescope} (\hst) and ALMA for a panchromatic view of SPT0311$-$58. This allows us to constrain the physical properties of the stellar component of SPT0311$-$58 by bridging the dusty and UV counterparts of this incredible system.

The outline of this paper is as follows: in \S\ref{sec:data} we describe the new and supplementary data utilized in this study, and in \S\ref{sec:methods} we describe our analysis, including profile-fitting photometry and spectral energy distribution (SED) modeling. In \S\ref{sec:results} we present the results and discuss their implications for the evolution of SPT0311$-$58. Finally, in \S\ref{sec:conclusions} we summarize our findings. We assume a flat $\Lambda$CDM cosmology with $\Omega_{m,0}=0.310$ and $H_0=67.7$\,km\,s$^{-1}$\,Mpc$^{-1}$ \citep{2020A&A...641A...6P}. At $z=6.9$, an angular size of $1^{\prime\prime}$ corresponds to a proper physical size of $\sim5.4$\,kpc.

\section{Data} \label{sec:data}

\subsection{\jwst\ NIRCam \& MIRI Imaging}

As part of the \jwst Cycle 1 program GO-1791 (PI: J. Spilker), we obtained imaging data from the NIRCam and MIRI instruments. The NIRCam images were taken in the F200W, F300M, F335M, and F444W filters with exposure times of 7945.2\,s, 2920.4\,s, 2920.4\,s, and 2104.4\,s, respectively. They cover the rest-frame UV and optical emission at $z\sim7$. The INTRAMODULE dither pattern was used with four dithers. Following the standard reduction pipeline \citep{bagley_ceers_2023}, the NIRCam images were processed using the \jwst Calibration Reference Data System (CRDS) version 11.17.14. The pixel scales of the final images were then resampled to $0.03^{\prime\prime}$/pixel.

The MIRI observations were obtained in the F560W, F770W, F1280W, and F1800W, covering the rest-frame near- and mid-IR. Each filter had an exposure time of 1110.0\,s. The default CYCLING dither pattern with four dithers was used. We combined our F560W and F770W images with the MIRI imaging obtained during the GTO-1264 Cycle 1 program (PI: L. Colina). This program obtained F560W and F770W imaging simultaneously with their proposed MRS observations. We did not use the MRS observations here. F560W was carried out in two 2-point dither observations with exposure times of 1859.3\,s and 1859.3\,s, while F770W was obtained in a single 4-point dither with an exposure time of 7448.2\,s.

We reduced the MIRI images by primarily following the workflow outlined by the MIRI observation and TEMPLATES teams \citep{rigby_templates_25} using CRDS version 11.18.1. We followed a recommended method from the Space Telescope Science Institute to mitigate the striping artifacts present in all \jwst data. We constructed the striping pattern from the median-stacked calibrated images, then subtracted this model pattern from each stage-2 image of the pipeline. This removed the artifacts, but caused some localized over-subtraction of extended emission around bright neighboring sources. Given the small number of dithers, bright, extended sources can leave localized imprints in the median stack, causing over-subtraction in their immediate vicinity. However, this is not an issue for this analysis because the core region of SPT0311$-$58 is not one of those impacted sources, which we confirmed upon visual inspection of individual maps and the median-stacked stripe model. The pixel scale was then resampled to 0.06$''$/pixel. 

To align the MIRI images, we first generated a star catalog using \texttt{Source Extractor} on the ancillary HST WFC3 F160W imaging, which were aligned to the Gaia DR3 catalog (see later discussion in \S\ref{subsec:suppdata}). Using the \texttt{SourceFinder} routine from \texttt{photutils}, we set a detection threshold of 40 times the background root-mean-square (RMS) value, which found 450 sources for our reference catalog. We then generated a source catalog for each MIRI image by setting a detection threshold of 2.5 times the background RMS value. After cross-referencing these catalogs with our F160W reference catalog, we matched between 50--100 sources per MIRI image, yielding an RMS of the final astrometric solution of about 0.05--0.1$''$ in each band.

\subsection{Ancillary Data} \label{subsec:suppdata}

\begin{figure*}[ht!]
    \centering
    \includegraphics[width=\linewidth]{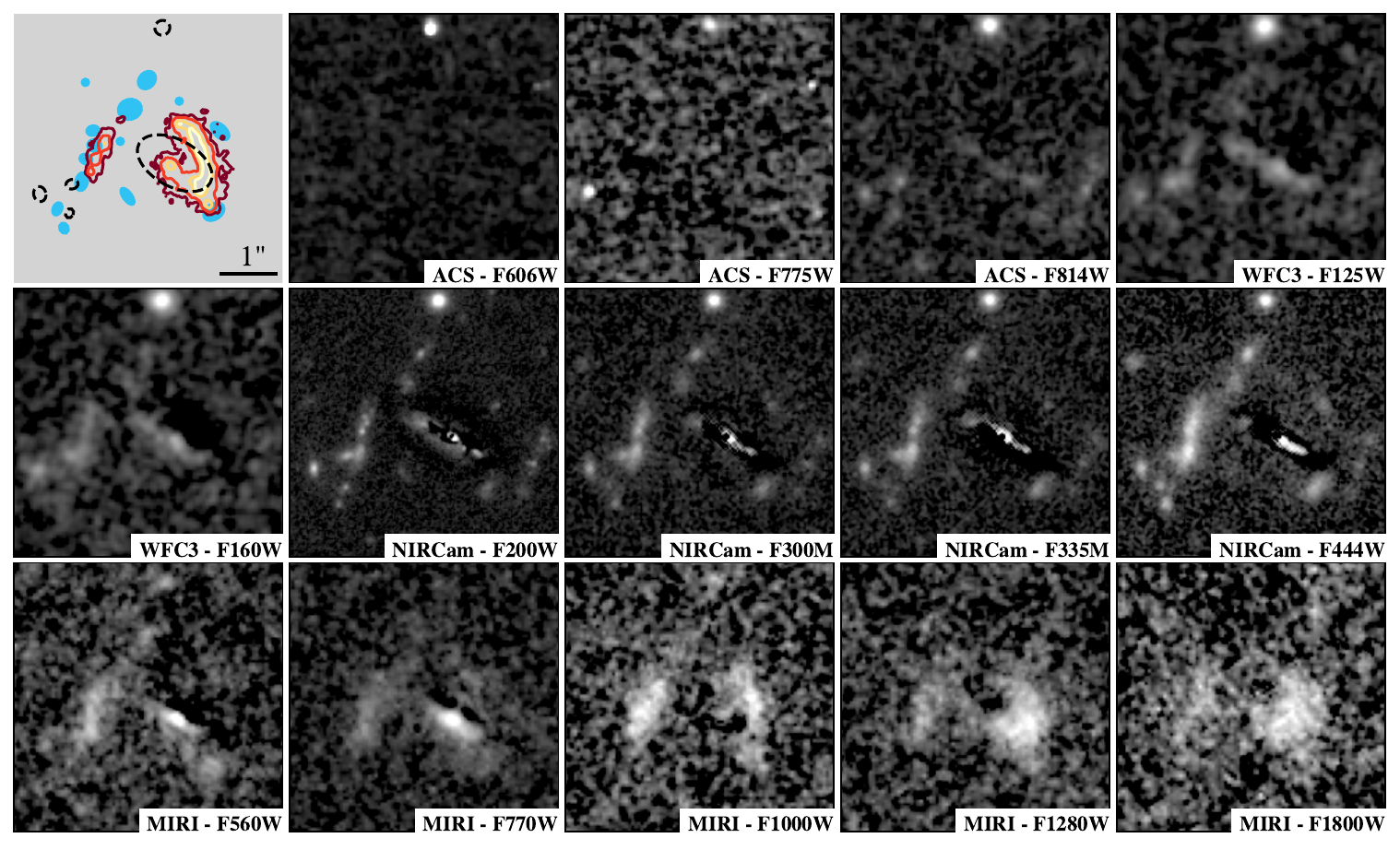}
    \caption{\hst, NIRCam, and MIRI $5^{\prime\prime}\times5^{\prime\prime}$ cutouts of the SPT0311$-$58 system for visualization, with imaging filters as indicated. Each image is displayed using a square root normalization ranging from the 25th to 99.9th percentile values. The top left panel indicates locations of named sources from Figure \ref{fig:rgb} in blue along with the same ALMA contours overlaid. Foreground sources are denoted by the black dashed lines. The foreground lensing galaxy, which is visible in all filters, has been removed using the two-component profile result discussed in \S \ref{subsec:profilephot}. The least obscured regions of SPT0311$-$58E are identifiable in the \hst/WFC3 imaging (rest-frame $\sim1800$\,\AA), while the bulk of SPT0311$-$58W is only visible in the MIRI images above $\sim1\,\mu$m.} 
    \label{fig:thumbnails}
\end{figure*}

We augmented the \jwst data with other imaging to expand our wavelength coverage and constrain the SED modeling. We use Band 6 ALMA data (project IDs 2016.1.01293.S and 2017.1.01423.S) reported by \citet{spilker_chaotic_2022}, hereafter \citetalias{spilker_chaotic_2022}, to provide coverage of the 160\,$\mu$m dust continuum in our SED fitting. \citetalias{spilker_chaotic_2022} identified the following 12 dusty clumps distributed between SPT0311$-$58E and W in either the 160\,$\mu$m dust or the 158\,$\mu$m [C{\sc ii}] maps:

\begin{itemize}
\item[\smolbullet] Four clumps in SPT0311$-$58E (E1, E2, E3, E4).

\item[\smolbullet] Eight clumps in SPT0311$-$58W (W1, W2, W3, W4, W5, W6, W7, W8). We treat W1-W7 as a single object, as discussed later in \S\ref{subsec:uv_and_dusty_sources}.
\end{itemize}

Throughout this work, ``clumps" refers specifically to connected regions that fall within SPT0311-58E and W. We adopt the naming scheme from \citetalias{spilker_chaotic_2022} for the dust obscured clumps. It is uncertain at present what these clumps represent. With sizes $\gtrsim1.6$\,kpc, they are too large to be star clusters or Giant Molecular Clouds. Based on the results from \citetalias{spilker_chaotic_2022}, the possibility of these clumps being fragments of larger disk-like structures or caused by interactions between SPT0311$-$58E and W cannot be ruled out. 

Photometric data from \hst previously reported in \citetalias{marrone_galaxy_2018} covers the rest-frame UV radiation and helps with constraining the Lyman-break. This includes five wide-band images: F606W, F775W, and F814W using ACS and F125W and F160W using WFC3 (PID 14740; PI: D. Marrone, PID 16237; PI: S. Chapman). Since the F606W and F775W filters are fairly shallow observations, we opted for the single combined mosaic from the Hubble Advanced Products directly. We utilized the \hst reduction pipeline for the rest of the \hst filters (F814W, F125W, and F160W). A few exposures were severely impacted by the failure of the Fine Guidance sensor during observation and were discarded. By this stage in the reduction, we found that not masking hot pixels led to clusters of bright pixels repeated across the images. While not the standard approach, we found that keeping saturated pixels and masking hot and warm pixels sufficiently reduced these artifacts.

As for astrometric alignment, we first aligned the F814W exposures to the Gaia DR3 catalog using \texttt{tweakreg}. The typical RMS of the final astrometric solutions were $\sim0.02''$. As we did with the \jwst imaging, we then used \texttt{Source Extractor} on the astrometry-corrected F814W image to generate a reference catalog to align F125W and F160W. The final astrometric solution for each filter was also $\sim0.02''$.

NIRSpec observations of the core of SPT0311$-$58 presented by \citet{arribas_ga-nifs_2024}, hereafter \citetalias{arribas_ga-nifs_2024}, identified rest-frame UV-dominant clumps within SPT0311$-$58E as well as additional nearby sources. Of these nearby sources, \citetalias{arribas_ga-nifs_2024} found that nearly all of them are gravitationally bound and confirmed these sources as members of this system. We use results reported by \citetalias{arribas_ga-nifs_2024} using the NIRSpec IFU data from the GA-NIFS GTO program (PID 1264; PI: L. Colina). With the exception of SPT0311$-$58W, which was too faint to be detected in their NIRSpec data, they identified the following sources within their field of view:

\begin{itemize}
\item[\smolbullet] Four low-$z$ sources (the lens, lz1, lz2, lz3).

\item[\smolbullet] Four clumps within SPT0311$-$58E (ES, ENa, ENb, ENc).

\item[\smolbullet] Ten spectroscopically-confirmed $z\sim6.9$ sources. Three were identified based on UV-continuum (C1, C2, C3), while seven were identified by their \oiii emission and classified as line emitters (L1, L2, L3, L4, L5, L6, L7). L7 lies along the same line-of-sight as ENa and was only distinguishable kinematically. We were unable to include this source in our analysis.

\end{itemize}

We adopt the naming scheme from \citetalias{arribas_ga-nifs_2024} for the unobscured sources for consistency and use their reported coordinates as starting points in our profile fitting model (see \S\ref{subsec:profilephot}). Their \halpha fluxes and approximate equivalent widths (EW) were used in our \textsc{CIGALE} SED fitting to help constrain ages and nebular emission strengths (see \S\ref{subsec:sed}). 

Lastly, in addition to the supplementary MIRI observations mentioned previously, we also use the F1000W imaging obtained during the GTO-1264 program (PI: L. Colina), which was previously reported in \citet{alvarez-marquez_mirijwst_2023}, hereafter \citetalias{alvarez-marquez_mirijwst_2023}. This deep F1000W image is more sensitive to SPT0311$-$58W than our MIRI imaging (see Figure \ref{fig:thumbnails}) and adds another photometric constraint for our SED fitting. We also compare our SED fitting models with the photometry presented in \citetalias{alvarez-marquez_mirijwst_2023} as a consistency check.

Known sources have been mapped out in Figure \ref{fig:rgb}, along with a potential new source just south of SPT0311$-$58E (labeled ``S'' for convenience). Cutouts of the \hst and \jwst images can be found in Figure \ref{fig:thumbnails}. The data presented in this article were obtained from the Mikulski Archive for Space Telescopes (MAST) at the Space Telescope Science Institute. These observations can be accessed via \dataset[doi:10.17909/wsfs-mq19]{https://doi.org/10.17909/wsfs-mq19}.

\section{Methods} \label{sec:methods}

Within just a $5^{\prime\prime}$ field of view of the core of SPT0311$-$58, several dozen sources are identifiable, including clumps and bound members part of SPT0311$-$58 itself, the foreground lens galaxy, and a few low-$z$ interlopers. Analyzing such a complicated and crowded field must be handled with care. For instance, we have divided our source detection to identify low- and high-obscuration sources separately. More details are provided in section \S\ref{subsec:uv_and_dusty_sources}. Additionally, we have opted to use profile-fitting photometry since standard aperture photometry is insufficient for this work. Aperture photometry requires matching the Point Spread Function (PSF) of \emph{all} of the bands to the lowest resolution image, sacrificing information gained from the higher resolution data. This also requires an aperture correction to account for the lost light due to the PSF. Profile fitting takes each image's resolution into account and does not require these steps. It also allows us to perform ``forced photometry'', meaning we can measure photometry on previously identified sources by assuming a fixed position across all photometric bands.

\subsection{Source Detection} \label{subsec:sourcedet}

To identify and characterize the emission from the individual clumps within SPT0311$-$58 and nearby sources, we employed \texttt{galight} \citep{ding_galaxy_2021}, a \textsc{Python}-based package that combines the source detection capabilities of \texttt{photutils} \citep{bradley_photutils_2016} and the profile-fitting photometry of the lensing code \texttt{lenstronomy} \citep{birrer2018lenstronomymultipurposegravitationallens, 2021JOSS....6.3283B}.

Given that the clumps vary greatly in dust composition, we cannot rely on a single photometric band or wavelength regime for source detection. As an example, the dusty galaxy SPT0311$-$58W does not appear in NIRCam nor is it resolved in our MIRI imaging. Therefore, we adopt a tailored approach to source detection, distinguishing between the UV-dominant and dusty sources, which we outline below.

\begin{figure*}[p]
    \centering
    \includegraphics[width=\linewidth]{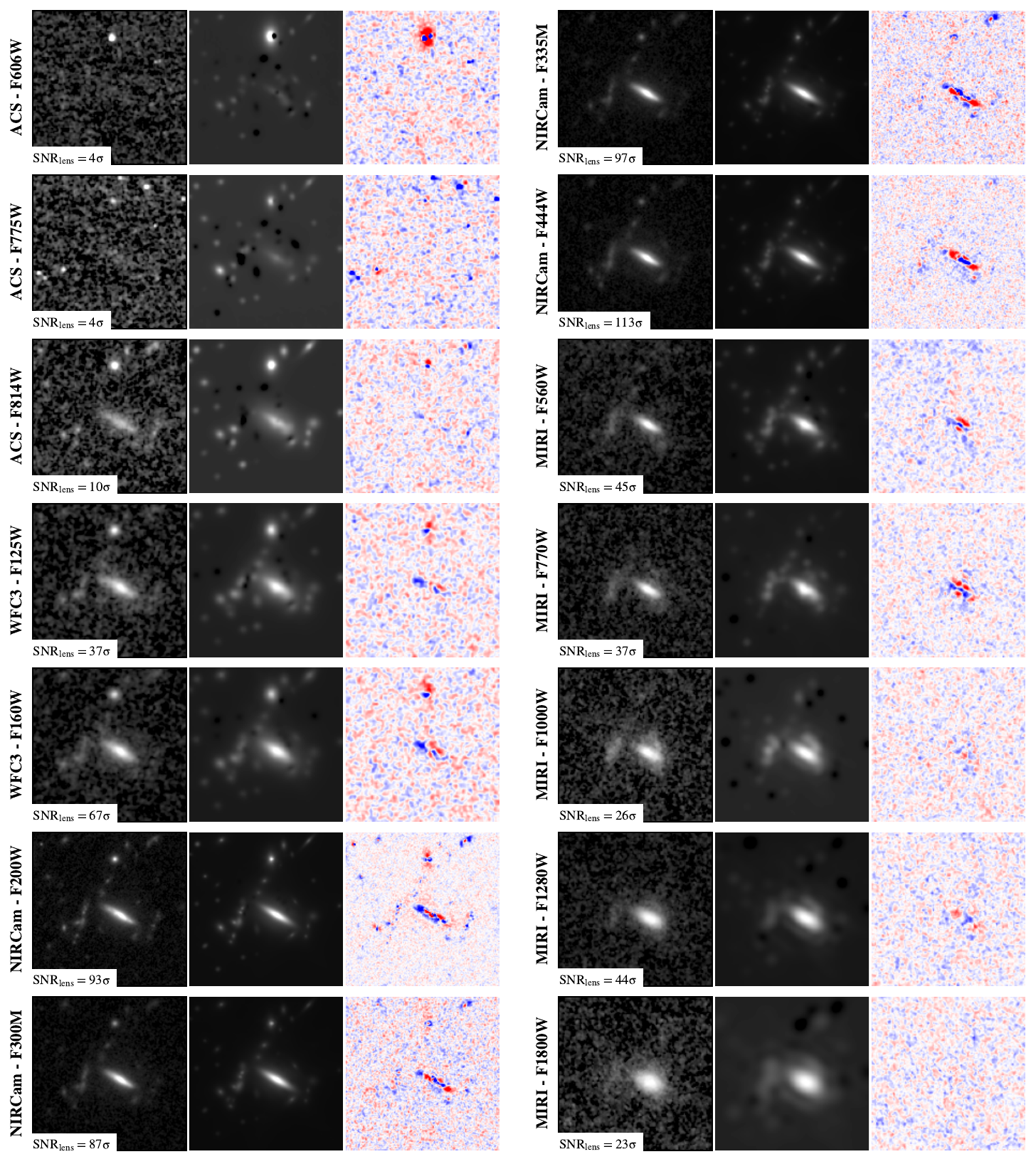}
    \caption{$7^{\prime\prime}\times7^{\prime\prime}$ cutouts for all 14 photometric filters showing the results of our profile-fitting photometry models using \texttt{galight}. Ordered by wavelength coverage, we show the data, model, and normalized residuals. The data and model columns are displayed using a square root normalization with color bounds set by the 25th and 99.9th percentile values. The color bar stretch for the residuals ranges from $\pm5\sigma$. Our profile-fitting model represents the data well with a median reduced $\chi^2$ of $0.92$. The peak observed pixel signal-to-noise ratios of the lens has been listed in the data panel.}   
    \label{fig:galight_models}
\end{figure*}

\subsubsection{Handling Low vs High Obscuration} \label{subsec:uv_and_dusty_sources}

Within the NIRCam imaging alone, we do not identify all of the NIRSpec-identified clumps nor any of the ALMA clumps. To detect the less dust-obscured sources, we construct a detection image by PSF-matching the four NIRCam exposures to the F444W resolution, then median-stacking them with inverse-variance weighting. The \texttt{webbpsf} software was used to simulate the PSF for each exposure and generate a transformation convolution kernel \citep{perrin_updated_2014}. The source detection was then performed on the PSF-matched detection image with a detection threshold of 3$\sigma$ above background. Comparing our detected clumps with the clumps found by \citetalias{arribas_ga-nifs_2024} with NIRSpec, it was evident that not all of the sources were detected or deblended. Manual adjustments were therefore made to account for all of the clumps identified by \citetalias{arribas_ga-nifs_2024}, as discussed later in \S\ref{subsec:profilephot}. For sources not previously identified by \citetalias{spilker_chaotic_2022} in the 160 micron ALMA image, upper flux limits were estimated by randomly placing 500 of the corresponding circularized aperture outside of the central 8$''$ and taking the median value.

Due to the high level of dust obscuration in SPT0311$-$58W (and to a lesser degree, SPT0311$-$58E), the clumps identified with ALMA are not detected in the NIRCam images. Therefore, a secondary source detection was performed on the 160\,$\mu$m dust continuum image from ALMA using a 6$\sigma$ detection threshold. Ten sources (plus the counter image produced by the lens) were detected, with some of the sources reported in \citetalias{spilker_chaotic_2022} remaining undetected (W7) or blended (W2 and W3). We only recover SPT0311$-$58W in our low-resolution, long-wavelength MIRI bands, and lack visual evidence in the rest frame optical bands. Additionally, the foreground lensing galaxy produces a counter image of SPT0311$-$58W. Attributing light from this counter image to individual clumps in SPT0311$-$58W requires a careful and robust lensing analysis that is beyond the scope of this work. Due to these complications, we chose to sum the fluxes of W1--W7 and treat them as a singular source. Given this decision, we did not make any further adjustments to account for the discrepancies mentioned with W2, W3, and W7.

\subsection{Profile-Fitting Photometry} \label{subsec:profilephot}

We opted for profile-fitting photometry using \texttt{galight}. This method fits a seven-parameter (two component position and ellipticity, $n_{\text{S}\acute{\text{e}}\text{rsic}}$, $R_{\text{S}\acute{\text{e}}\text{rsic}}$, and amplitude) elliptical S\'{e}rsic profile to each of the detected sources. All sources detected within a desired region are modeled simultaneously. Therefore, profile-fitting photometry eliminates the need to subtract contaminating sources prior to modeling, such as the foreground lens.

While our primary targets fall within the inner $5^{\prime\prime}$ of SPT0311$-$58, we model the inner $7^{\prime\prime}$ for a conservative edge buffer. Using the stacked NIRCam image as the detection image, we identify 39 sources above our 3$\sigma$ detection threshold in addition to two point sources above 30$\sigma$, which translates to approximately 280 free parameters. From here, \texttt{galight} estimates starting values based on the pixel distribution of each source. We enforced a minimum source radius of 0.3$''$ ($\sim$\,1.6\,kpc), so any identified objects below this were fixed to circular sources of this radius with a Gaussian profile instead to help reduce the dimensionality of our fit. Center positions of the profile were allowed to search within the surrounding 4\,$\times$\,4 pixels of the starting position. At this point, the following manual adjustments were made to account for the regions identified by \citetalias{arribas_ga-nifs_2024}, using their reported coordinates as starting positions. 
\begin{itemize}
    \item[\smolbullet] L2, L3, and L4 were not detected, therefore circular sources matching our minimum radius criteria (0.3$''$) were added. As a result, fluxes for these sources are mostly upper limits.
    \item[\smolbullet] ENa, ENb, and ENc did not have a high enough contrast to deblend, so this was done manually.
    \item[\smolbullet] ES and lz3 were also separated manually.
\end{itemize}

Additionally, we found that the lensing galaxy was not well represented as a single S\'{e}rsic profile. The residuals in our initial fits showed low-level diffuse emission along the outskirts of the lensing galaxy, possibly due to a recent disturbance. This diffuse emission was contaminating compact SPT0311$-$58 sources, and the model preferred extending these compact sources in an attempt to account for this diffuse component. We therefore modeled the foreground lensing galaxy as two S\'{e}rsic profiles, one for the prominent disk-like structure and one for the larger diffuse emission. We enforced these two profiles to share the same center location and that the disk-like component be smaller and more elliptical than the diffuse component. Indeed, upon visual inspection, the addition of this diffuse component both sufficiently accounted for the diffuse residuals around the lensing galaxy and allowed the smaller nearby sources to retain their compact shapes. Previous studies have also found that distant galaxies observed by \jwst are better represented as two components, either for bulge-disk \citep{tang_unveiling_2025} or thin and thick disk \citep{tsukui_emergence_2025} decomposition.

From our results in Figure \ref{fig:galight_models}, our two-component model still leaves $\sim5\sigma$ residuals at the center of the lensing galaxy. This is a significant improvement from the peak $10$--$113\sigma$ lens signal prior to modelling (see Figure \ref{fig:galight_models} for details). This may suggest that the lensing galaxy has more complex structure, such as spiral arms, that are not well represented by our two S\'{e}rsic profiles. We tested models to replicate the detail in the lensing galaxy by adding additional S\'{e}rsic components. These did little to reduce the peak residuals, and additionally could coincide with lensed light from SPT0311$-$58W. As discussed previously in \S \ref{subsec:uv_and_dusty_sources}, a detailed lensing analysis is necessary to confidently justify these additional lens light components.

A coarse-grain particle swarm optimization (PSO) was run first to quickly narrow down the parameter space, the results of which were then used as initial estimates for a fine-grain Markov Chain Monte Carlo (MCMC) fit. Once the detection fit was completed, we then used the \texttt{MultiBandImageReconstruction()} function from \texttt{lenstronomy} to fit all 14 of our images, which performs a weighted least squares multi-band fit to fit only the amplitudes of each source while keeping the remaining profile parameters fixed. This was done on the original images since their resolution was taken into account during the multi-band fitting procedure. 

We assessed our astrometric alignment using the residuals from our model. Since our model was built from the NIRCam detection image, any offset would be corrected to align with the NIRCam observations. We found a slight tangential offset with MIRI bands F560W, F770W, F1280W, and F1800W. \texttt{lenstronomy} has built-in alignment capabilities on sub-pixel scales. We utilized this function to align each of the four MIRI bands to the NIRCam detection image by only allowing for shifts in right ascension (RA) and declination (Dec.); no rotational alignment was included. We then applied the average offset in both RA and Dec. to correct the astrometry. Final model results can be found in Figure \ref{fig:galight_models}.

Based on the proximity of the dusty and non-dusty clumps, as seen in Figure \ref{fig:rgb}, we have reason to believe that some of these previously-identified clumps are actually components of the same clump with varying degrees of obscuration. Therefore, we have asserted the following pairs as single clumps for SED fitting purposes: E1+C1, E2+ENb, E3+ENa, E4+ES, and W8+C3. Profile fitting was carried out on the individual clumps, then summed accordingly prior to SED fitting. 

To test the robustness of our profile-fitting photometry, we compare our photometric measurements in F1000W to the original aperture photometry measured by \citetalias{alvarez-marquez_mirijwst_2023}. They show the overlap in their identified clumps to the clumps identified in \citetalias{spilker_chaotic_2022}, and we use these overlaps for comparing photometry measurements. Given our own combination of clumps, this means we compare their EA to our E2+ENb, EB+EC to E3+ENa, EE+ED to E4+ES, and their integrated W to our W1-7. We find that all of these pairings agree within $1\sigma$ uncertainty, showing that we recover sufficiently accurate photometry on clumpy and global scales.

\subsection{SED Fitting} \label{subsec:sed}

\begin{figure*}
    \centering
    \includegraphics[width=\linewidth]{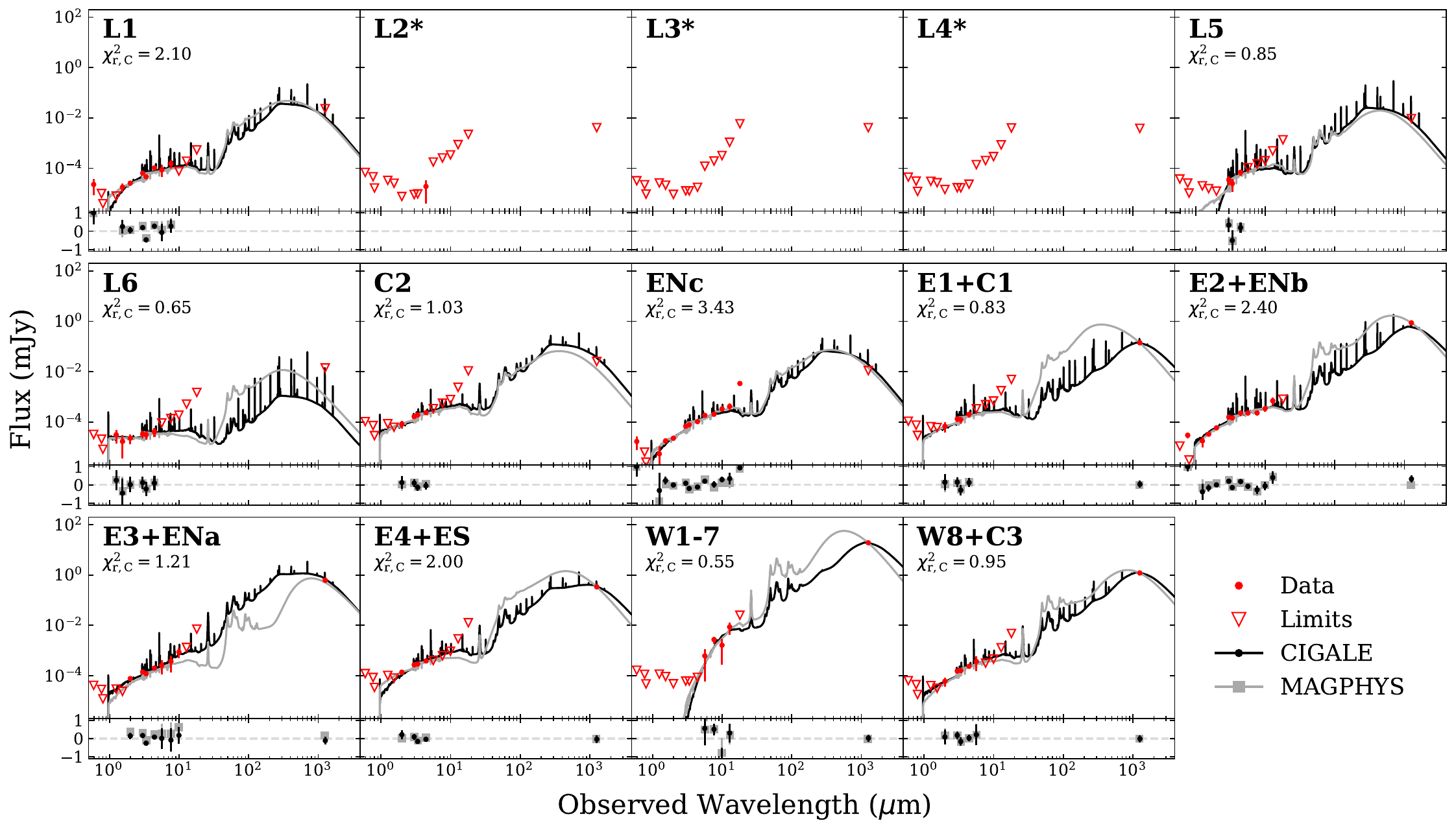}
    \caption{
    Resulting best-fit SEDs from \textsc{cigale} (black lines) and \textsc{magphys} (grey lines) for the SPT0311$-$58 members considered in this work. We omit L2, L3, and L4, as they are detected in at most one photometric band. We find reasonably good agreement between the best-fit SEDs found from the \textsc{cigale} and \textsc{magphys} codes for most of the SPT0311$-$58 clumps despite their different assumptions and treatment of emission lines. Photometry from our joint multi-band, multi-component forward modeling procedure is shown in red circles. Upper limits (1$\sigma$) are shown with downward arrows. The lower panel of each plot indicates the fractional residuals (data minus model / data) in black circles (for \textsc{cigale} residuals) and grey squares (\textsc{magphys}). No correction for magnification has been applied.
    }
    \label{fig:sed}
\end{figure*}

To estimate key clump and global properties of SPT0311$-$58, we perform SED modeling on the multi-band photometry in this section. Our SED fits were tested against previous studies of SPT0311$-$58, particularly \citetalias{alvarez-marquez_mirijwst_2023} and \citetalias{marrone_galaxy_2018}. We employed two different SED fitting codes, \textsc{cigale} and \textsc{magphys}, to cross-validate our results. More details about the fitting are provided in this section.

Since L2, L3, and L4 were detected in at most one imaging band, the SED fitting is unreliable. For stellar mass, we estimate upper mass limits from the $\langle M_{\star}/L_{\rm F444W} \rangle$ of the remaining line-emitters (L1, L5, and L6), where $M_{\star}$ is the SED derived stellar mass. $\langle M_{\star}/L_{\rm F444W} \rangle$ is calculated for both \textsc{cigale} and \textsc{magphys}. We then multiply $\langle M_{\star}/L_{\rm F444W} \rangle$ by the $3\sigma$ upper limit of the F444W flux. For SFR, we report the \halpha derived SFRs from \citetalias{arribas_ga-nifs_2024} when available.

\subsubsection{\textsc{cigale}} \label{subsubsec:cigale}

\textsc{cigale} \citep{boquien_cigale_2019} is commonly applied to the photometry of DSFGs, allowing for the simultaneous modeling of UV--radio emission using an energy balance approach. It offers a flexible set of parametric star-formation histories (SFHs), which is well-suited for our analysis. We use \textsc{cigale} version 2025.0 and test several SFH parameterizations to probe the sensitivity of our data to SED model priors. As our fiducial model, we start with a simple constant SFH where we assume stellar ages ranging from 1--750\,Myr with grid values of 1, 10, 25, 50, 75, and 100--750\,Myr in 50\,Myr increments. We assume solar metallicity using the stellar population models from \citet{bruzual_stellar_2003} and a Chabrier initial mass function \citep{chabrier_galactic_2003}. For our tests, we run the following SFH models using the same assumptions unless otherwise specified:

\begin{itemize}
    \item[\smolbullet] Constant SFH, but allowing the metallicity to vary with grid values of 0.0004, 0.004, 0.008, 0.02, and 0.05 (approximately 0.02--2\,Z$_\odot$).
    \item[\smolbullet] Constant SFH, but allowing an optional constant burst. The main stellar population age grid values are 25, 50, 75, and 100--750\,Myr in 50\,Myr increments, and the burst ages are 1, 10, and 25\,Myr. The burst mass fraction grid values are 0, 0.01, 0.1, 0.5, and 0.99.
    \item[\smolbullet] Delayed exponential with no burst. The allowed $e$-folding time values were 1, 10, 100, 1000, 10000, and 13000\,Myr.
    \item[\smolbullet] Delayed exponential with an optional constant burst. The main stellar population age grid values are 25, 50, 75, and 100--750\,Myr in 50\,Myr increments, and the burst ages are 1, 10, and 25\,Myr. The burst mass fraction grid values are 0, 0.01, 0.1, 0.5, and 0.99.
\end{itemize}

The nebular, dust attenuation, and dust emission module parameters are similar to those presented in \citetalias{alvarez-marquez_mirijwst_2023}. For the nebular emission, we assumed solar metallicity, an electron density of 100\,cm$^{-3}$, $\log(U)=-3$, and a Lyman escape fraction of zero. We employed the dust attenuation models from \citet{charlot_simple_2000} and used the double power law parameters outlined in Table 1 of \citet{lo_faro_characterizing_2017} to motivate our grid parameters. Note that the parameters defined by \citet{lo_faro_characterizing_2017} are different than the input parameters accepted by \textsc{cigale}, which we have accounted for. Thus, the $V$-band attenuation due to the ISM ($A_\text{V,ISM}$) ranges from 0.1--10.0 and the fraction of the ISM attenuation to the total ($A_\text{V,ISM}/(A_\text{V,BC}-A_\text{V,ISM})$) from 0.4--1.0. This allows the birth cloud (BC) attenuation to range from 0--15. The power law slopes for both the ISM and BCs range from $-$3 to 0. 

The dust and IR emission models from \citet{draine_infrared_2007} were used. Since we do not have any PAH coverage, we kept the PAH fraction set to the default value of 2.5. The minimum scaling factor for the radiation field was allowed to vary from 1--25, and $\gamma$ (the fraction of dust heated by stars) from 0.02--0.5, all while keeping the maximum radiation field set to the default value of $1\times10^6$. Previous studies have not found evidence for an AGN in SPT0311$-$58, so we have not included an AGN component in our fitting. For sources with \halpha EW estimates, \textsc{cigale} measures the rest-frame EW of each SED model using a central bandwidth as well as a red and blue bandwidth to estimate the continuum. We allowed the bandwidths to range from 650--652.5\,nm for the blue continuum, 655.5--657.0\,nm for the central peak, and 661--663.5\,nm for the red continuum. These were chosen to avoid contamination from nearby [NII] features. 

When comparing the best-fit SED results for our SFH parameterization tests, we find that the reduced $\chi^2$ changes from the constant SFH by $\sim0.055$ for the delayed+burst model and by $\lesssim0.015$ for the remaining three models. Portions of the SED where we lack photometric coverage show the greatest variation, but the overall quality of the fits does not depend significantly on our adopted SFH priors. For most of the regions we consider here, we find that stellar mass and SFR estimates for all tested SFHs are consistent with the constant SFH within $1\sigma$ uncertainty. We adopt the constant SFH SED model results, as others have previously done (\citetalias{alvarez-marquez_mirijwst_2023}).

We then applied our constant SFH SED model to the photometric data provided in \citetalias{alvarez-marquez_mirijwst_2023} to ensure consistency. Indeed, the results for age, stellar mass, SFR, and $A_\text{V,ISM}$ agree with \citetalias{alvarez-marquez_mirijwst_2023} to within $1\sigma$. By replicating their results and confirming that our model priors do not affect our stellar mass and SFR measurements, we can attribute differences in parameter estimates from SED modeling to differences in the data used, since our analysis samples more of the rest-frame UV than \citetalias{alvarez-marquez_mirijwst_2023}.

Our resulting best-fits for the constant SFH are shown in Figure \ref{fig:sed}. The SED models recover the 160\,$\mu$m dust continuum within $1\sigma$ for most of the sources with this detection. The UV end of the fits exhibits the greatest residuals, though most still fall within the $1\sigma$ uncertainties, and the majority of the fits have $\chi^2_r$\,$\lesssim$\,1.

\begin{figure}
    \centering
    \includegraphics[width=\linewidth]{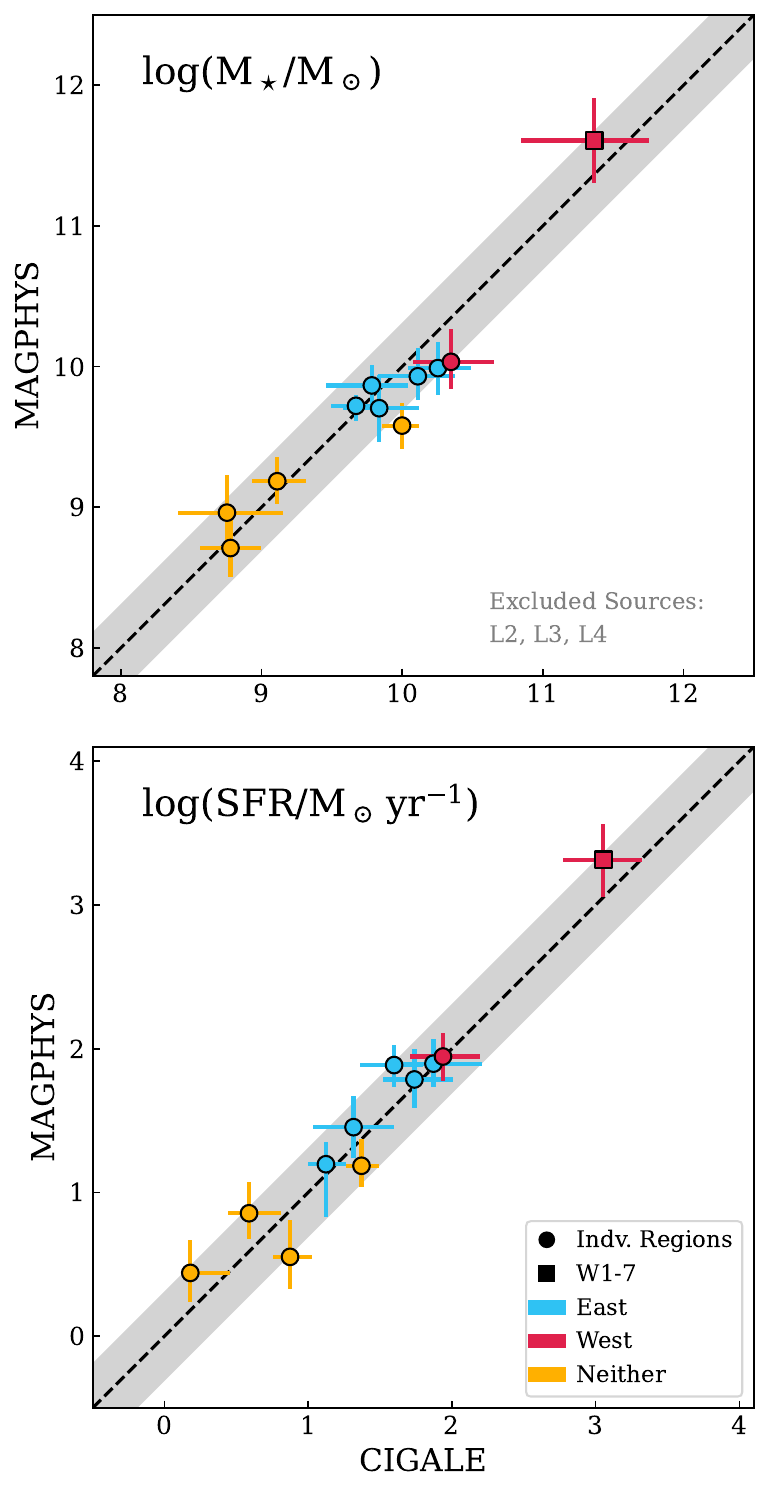}
    \caption{Comparing clump stellar masses and SFRs derived from \textsc{cigale} and \textsc{magphys}. The two SED codes mostly agree with each other within a factor of two (shaded region). The expected one-to-one relation is shown by the dashed line. L2, L3 and L4 had one or no photometric detections and were excluded from this comparison.}
    \label{fig:CvsM}
\end{figure}

\subsubsection{\textsc{magphys}} \label{subsubsec:magphys}

\textsc{magphys} is commonly applied to DSFGs because it handles dust attenuation and emission self-consistently by using an energy balance technique to model the SED of the sources from the UV/optical to the submillimetre/radio wavebands \citep[e.g.][]{da_cunha_alma_2015, miettinen17, dudzeviciute20, birkin21}. SFHs are modeled as continuous delayed exponentials with randomly drawn times for the peak of star formation and random bursts superimposed to model starbursts \citep{lee10}. Like \textsc{cigale}, \textsc{magphys} uses the stellar population models from \citet{bruzual_stellar_2003} and a Chabrier initial mass function \citep{chabrier_galactic_2003}. However, \textsc{magphys} allows the metallicity to vary between $Z$\,$=$\,0.02--2\,Z$_\odot$ rather than fixing $Z$\,$=$\,Z$_\odot$. We refer the reader to \citet{da_cunha_simple_2008, da_cunha_alma_2015} for a more comprehensive discussion of \textsc{magphys} and the energy balance technique, and to \cite{hunt2019} for a thorough comparison of the two codes.

The \textsc{magphys} best fits are also shown in Figure \ref{fig:sed}. Like \textsc{cigale}, the photometry is well-modeled, with $\chi^2_r$\,$\sim$\,1 in most cases. The two codes only differ significantly in the far-infrared, where we lack photometric coverage. We also note here that L3 and L4 are not fit with \textsc{magphys}, as they only have photometric upper limits, and \textsc{magphys} does not take emission line fluxes or EWs as an input. 

In Figure \ref{fig:CvsM}, we compare the physical properties derived using the two codes. For stellar mass, SFR, and total dust extinction, we find that they are consistent to within $1\sigma$ uncertainties for most of the sources we were able to obtain SED fits for. As for stellar ages, we find that \textsc{magphys} produces ages that are on average three times lower than the ages produced by \textsc{cigale}. Throughout this paper, we focus on the stellar properties we can consistently constrain, such as the stellar mass and SFR.

\section{Results and Discussion} \label{sec:results}

As one of the most extreme systems of DSFGs at high redshift, SPT0311$-$58 offers critical insights into the processes driving galaxy assembly in the early Universe. Although extensively observed with ALMA, the launch of \jwst has provided new opportunities for in-depth analysis. Having outlined the data and methods used in this study (see Sections \S\ref{sec:data} and \S\ref{sec:methods}), we now turn to a discussion of the key results and their implications.

\subsection{Visual Observations}

By combining UV and dust coverage, we have assembled a more complete picture of this system. We observed a dust gradient across SPT0311$-$58E, since the ALMA flux is concentrated on the western side of SPT0311$-$58E, confirming what previous work has found \citepalias{marrone_galaxy_2018, alvarez-marquez_mirijwst_2023, arribas_ga-nifs_2024}. The gradient is most prominent at the northern end, while the southern end has more overlapping UV and dust content. 

Our F200W image covers the rest-frame UV and is the most detailed picture of the unobscured stellar component of SPT0311$-$58 to date. As seen in Figure \ref{fig:rgb}, the F200W panel clearly resolves the clumps previously identified in NIRSpec by \citetalias{arribas_ga-nifs_2024}. The potential new member we have called S can also be seen in Figure \ref{fig:rgb} just south of SPT0311$-$58E. S has similar color properties to the less obscured sources, particularly the nearby line-emitter L5. We do not detect S in F606W or F775W, which would imply that it resides at $z>6$. However, S is quite small and could simply be faint and unresolved in those \hst filters. We used \textsc{cigale} to estimate the photometric redshift for S using a grid of 50 evenly spaced redshifts ranging from 1--10. With 99.7\% confidence, we found a broad redshift constraint of $z=1.75^{+3.25}_{-0.72}$. Therefore, S is likely not part of the larger SPT0311$-$58 system. However, it falls just outside the NIRSpec IFU footprint and therefore lacks spectroscopic coverage. Follow-up spectroscopic observations are necessary to confirm the redshift of S.

\subsection{Star-Forming Main Sequence}

\begin{figure}[t!]
    \centering
    \includegraphics[width=\linewidth]{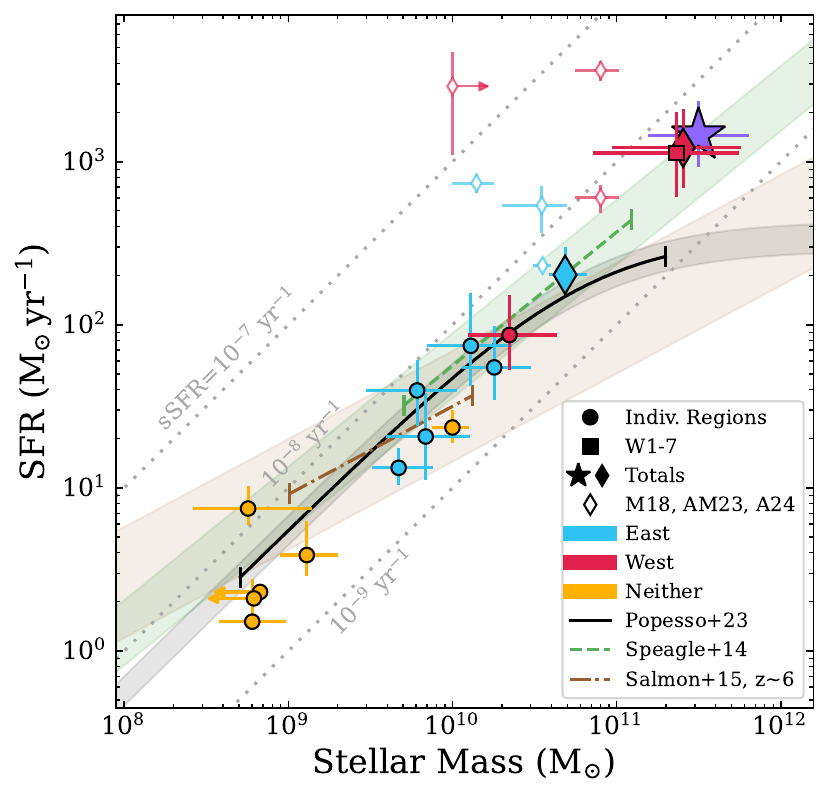}
    \caption{
    Comparing individual regions (circles) of SPT0311$-$58 as well as the overall system with the SFMS at $z\sim7$. Individual regions include the clumps associated with SPT0311$-$58E (blue) or W (red), as well as the five line-emitters and one continuum source, C2 (denoted ``Neither'', orange). L2 does not have a a reliable SFR from SED fitting nor \halpha, and thus is not shown. Most of the regions we identify in SPT0311$-$58 are consistent with the typical star-forming galaxy at $z\sim7$, but the system overall lies well above the main sequence as a result of the extremely high SFR of SPT0311$-$58W. 
    Diamonds show the sum total mass and SFR of SPT0311$-$58E and W, while the total over the entire central region is shown as a large star. The small white filled diamonds are results from \citetalias{marrone_galaxy_2018}, \citetalias{alvarez-marquez_mirijwst_2023}, and \citetalias{arribas_ga-nifs_2024}. All values have been corrected for lensing magnification. We show the SFMS relations from \citet{popesso_main_2023}, \citet{salmon_relation_2015}, and \citet{speagle_highly_2014} along with their reported scatters (shaded regions) of 0.09, 0.34, and 0.2\,dex. The mass ranges for which the original authors report the SFMS relations as constrained are plotted between the vertical dashed lines. The scatter of each relation has been extrapolated over a broader mass range for ease of visual comparison.
    }
    \label{fig:sfms}
\end{figure}

\begin{table}[t]
\centering
\scriptsize 
\begin{tabular}{l|llll}
\hline\hline
 & \multicolumn{2}{c}{\footnotesize{\textsc{cigale}}}& \multicolumn{2}{c}{\footnotesize{\textsc{magphys}}}\\
\hline
Source & $M_\star$& SFR& $M_\star$&SFR\\
 & {($\times10^9$\,M$_\odot$)}& {(M$_\odot\,\text{yr}^{-1}$)}& {($\times10^9$\,M$_\odot$)}&{(M$_\odot\,\text{yr}^{-1}$)}\\
\hline
L1 & $1.3^{+0.72}_{-0.41}$ & $3.9^{+2.3}_{-1.0}$ & $1.5^{+0.66}_{-0.45}$ & $7.2^{+4.2}_{-2.3}$ \\[0.8ex]
L5 & $0.57^{+0.82}_{-0.3}$ & $7.5^{+2.8}_{-1.5}$ & $0.92^{+0.71}_{-0.36}$ & $3.6^{+2.7}_{-1.4}$ \\[0.8ex]
L6 & $0.6^{+0.36}_{-0.23}$ & $1.5^{+1.3}_{-0.11}$ & $0.51^{+0.32}_{-0.18}$ & $2.8^{+1.8}_{-0.98}$ \\[0.8ex]
C2 & $10^{+2.7}_{-2.5}$ & $23^{+6.5}_{-4.6}$ & $3.8^{+1.5}_{-1.1}$ & $15^{+7.6}_{-4.1}$ \\[0.8ex]
ENc & $4.7^{+2.9}_{-1.4}$ & $13^{+4.4}_{-2.9}$ & $5.3^{+0.78}_{-1.0}$ & $16^{+6.0}_{-8.8}$ \\[0.8ex]
E1+C1 & $6.9^{+5.9}_{-2.9}$ & $21^{+18}_{-9.5}$ & $5.1^{+3.4}_{-2.0}$ & $28^{+17}_{-11}$ \\[0.8ex]
E2+ENb & $6.1^{+4.5}_{-3.1}$ & $40^{+21}_{-16}$ & $7.4^{+2.6}_{-1.8}$ & $77^{+26}_{-21}$ \\[0.8ex]
E3+ENa & $13^{+10}_{-6.0}$ & $74^{+83}_{-32}$ & $8.5^{+4.6}_{-2.6}$ & $79^{+34}_{-23}$ \\[0.8ex]
E4+ES & $18^{+12}_{-6.5}$ & $55^{+44}_{-20}$ & $9.8^{+4.7}_{-3.3}$ & $61^{+35}_{-21}$ \\[0.8ex]
W1-7 & $229^{+320}_{-160}$ & $1100^{+890}_{-520}$ & $409^{+380}_{-200}$ & $2100^{+1500}_{-890}$ \\[0.8ex]
W8+C3 & $22^{+21}_{-9.9}$ & $87^{+65}_{-34}$ & $11^{+7.1}_{-3.7}$ & $88^{+36}_{-26}$ \\[0.8ex]
\hline
L2* & $<0.82$ & \nodata & $<0.82$ & \nodata \\[0.8ex]
L3* & $<0.67$ & $(2.3)$ & $<0.67$ & $(2.3)$ \\[0.8ex]
L4* & $<0.62$ & $(2.1)$ & $<0.62$ & $(2.1)$ \\[0.8ex]
\hline
W & $250^{+320}_{-160}$ & $1200^{+890}_{-520}$ & $420^{+380}_{-200}$ & $2200^{+1500}_{-890}$ \\[0.8ex]
E & $49^{+18}_{-9.9}$ & $200^{+98}_{-42}$ & $36^{+7.9}_{-5.1}$ & $260^{+58}_{-40}$ \\[0.8ex]
\hline
Total & $320^{+320}_{-160}$ & $1500^{+890}_{-530}$ & $459^{+380}_{-200}$ & $2400^{+1500}_{-890}$ \\[0.8ex]
\hline\hline
\end{tabular}
\caption{Stellar masses and SFRs from the SED clump fitting. Uncertainties are derived from the 16th and 84th percentiles. Limits have been reported for regions that have less than three photometric detections. For L2, L3, and L4, we report $3\sigma$ mass limits using $\langle M_{\star}/L_{\rm F444W}\rangle \approx 1.42\times10^{10}\,\text{M}_{\odot}/\mu\text{Jy}$ (the value found for the fits to L1, L5, and L6 in both \textsc{cigale} and \textsc{magphys}), and the SFR values from \citetalias{arribas_ga-nifs_2024} where available. The values for SPT0311$-$58E, W, and overall total are summed post-fitting. All values have been corrected for magnification using the individual magnifications reported in \citetalias{arribas_ga-nifs_2024}.}
\label{tab:props}
\end{table}

High-redshift DSFGs can exhibit incredibly high SFRs, but their sSFRs can vary greatly when compared to the SFMS \citep{daddi_multiwavelength_2007,michalowski_stellar_2012}. A challenge with determining sSFRs for DSFGs comes from the difficulty in robustly deriving stellar masses \citep[although see e.g.][]{hainline_slar_2011, michalowski_cosmic_2010} due to heavy dust obscuration and limited rest-frame optical data \citep{swinbank_rest-frame_2004}. Our new data from \jwst enable more reliable stellar mass estimates, allowing for better placement of SPT0311$-$58 relative to the SFMS, shedding light on whether these systems represent short-lived starburst phases or the high-mass end of normal star formation at early cosmic times.

In Figure \ref{fig:sfms}, we show how the individual clump and global properties of SPT0311$-$58 compare with typical star-forming galaxies just 800\,Myr after the Big Bang. \citet{popesso_main_2023} find that the high-mass turnover in the SFMS may exist out to $z\sim6$, and we show their SFMS relation evaluated at 800\,Myr (see their Equation 14 and Table 2). Building off of the work presented in \citet{speagle_highly_2014}, \citet{popesso_main_2023} expands their sample to cover a wider range of redshifts ($z=0-6$) and construct a broader mass-complete sample ($10^{8.7}-10^{11.3}$\,M$_\odot$). By expanding the low-mass coverage by an order of magnitude, \citet{popesso_main_2023} can more effectively fit the linear regime of the SFMS and detect high-mass deviations. While the SFMS presented in \citet{popesso_main_2023} has both the mass and redshift coverage suitable for this analysis, we acknowledge the ongoing discourse regarding the presence of a high-mass turnover in the early Universe. Therefore, we also show the SFMS relations from \citet{speagle_highly_2014} as well as \citet{salmon_relation_2015} in Figure \ref{fig:sfms}. The three relations do not all agree at high masses ($>10^{11}$\,M$_\odot$), which could be indicative of the increased scatter along the SFMS seen in the early Universe \citep{cole_ceers_2025}.

In terms of global properties, we find that SPT0311$-$58E has significantly lower stellar mass than SPT0311$-$58W ($M_\star$\,$=$\,4.9$^{+1.8}_{-0.99}\times10^{10}$\,M$_\odot$ and $M_\star$\,$=$\,$2.5^{+3.2}_{-1.6}\times10^{11}$\,M$_\odot$, respectively) and is forming stars at a much lower rate (SFR\,$=$\,$200^{+98}_{-42}$\,M$_\odot\,\text{yr}^{-1}$ and SFR\,$=$\,$1200^{+890}_{-520}$\,M$_\odot\,\text{yr}^{-1}$, respectively). SPT0311$-$58E agrees with all three SFMS relations and seems to represent a typical star-forming galaxy in the early universe. The interpretation of SPT0311$-$58W, however, depends on the SFMS. While it is believed to be experiencing a burst of star formation, this conclusion is uncertain given the scatter between the SFMS relations at the high-mass end. We also recall from our SFH tests that SED models with a constant SFH and those with a recent burst are equally consistent with the data available. When comparing the stellar properties of each region, we see that most of them behave as typical star-forming galaxies at this epoch. Stellar masses and SFRs for individual clumps derived from both \textsc{cigale} and \textsc{magphys} can be found in Table \ref{tab:props}.

The previous measurements for SPT0311$-$58E and W from \citetalias{marrone_galaxy_2018}, \citetalias{alvarez-marquez_mirijwst_2023}, and \citetalias{arribas_ga-nifs_2024} are also shown in Figure \ref{fig:sfms}. While these measurements agree to within a factor of two, we attribute the spread seen in stellar mass and SFR to differences in data and methods used to derive these quantities. \citetalias{marrone_galaxy_2018} relied on IR luminosity, \citetalias{alvarez-marquez_mirijwst_2023} used SED fitting which primarily used dust continuum, and \citetalias{arribas_ga-nifs_2024} used rest frame UV stellar continuum and emission. \citetalias{arribas_ga-nifs_2024} estimated their SFR using their \halpha measurement and followed the same extinction correction procedure using $A_V(\text{Pa}\alpha/\text{H}\alpha)$ as done in \citetalias{alvarez-marquez_mirijwst_2023}. While not as significant for SPT0311$-$58E, \citetalias{alvarez-marquez_mirijwst_2023} noted that their extinction-corrected SFR(\paalpha) only recovered $\sim10$\% of the IR-derived SFR from \citetalias{marrone_galaxy_2018} for SPT0311$-$58W. They noted other studies that also found that the SFR(\halpha), even after attenuation corrections, was consistently less than SFR(\paalpha) \citep{rieke_determining_2009, piqueras_lopez_vlt-sinfoni_2016}. \citetalias{arribas_ga-nifs_2024} found similar results, mentioning that their extinction corrected SFR(\halpha) values only recovered about a fifth of what \citetalias{marrone_galaxy_2018} had previously found. The differences described here emphasize the importance of a panchromatic approach when measuring stellar masses and SFRs of high-$z$ galaxies.

\subsection{Cosmological Extreme}

\begin{figure*}[ht!]
    \centering
    \includegraphics[width=\linewidth]{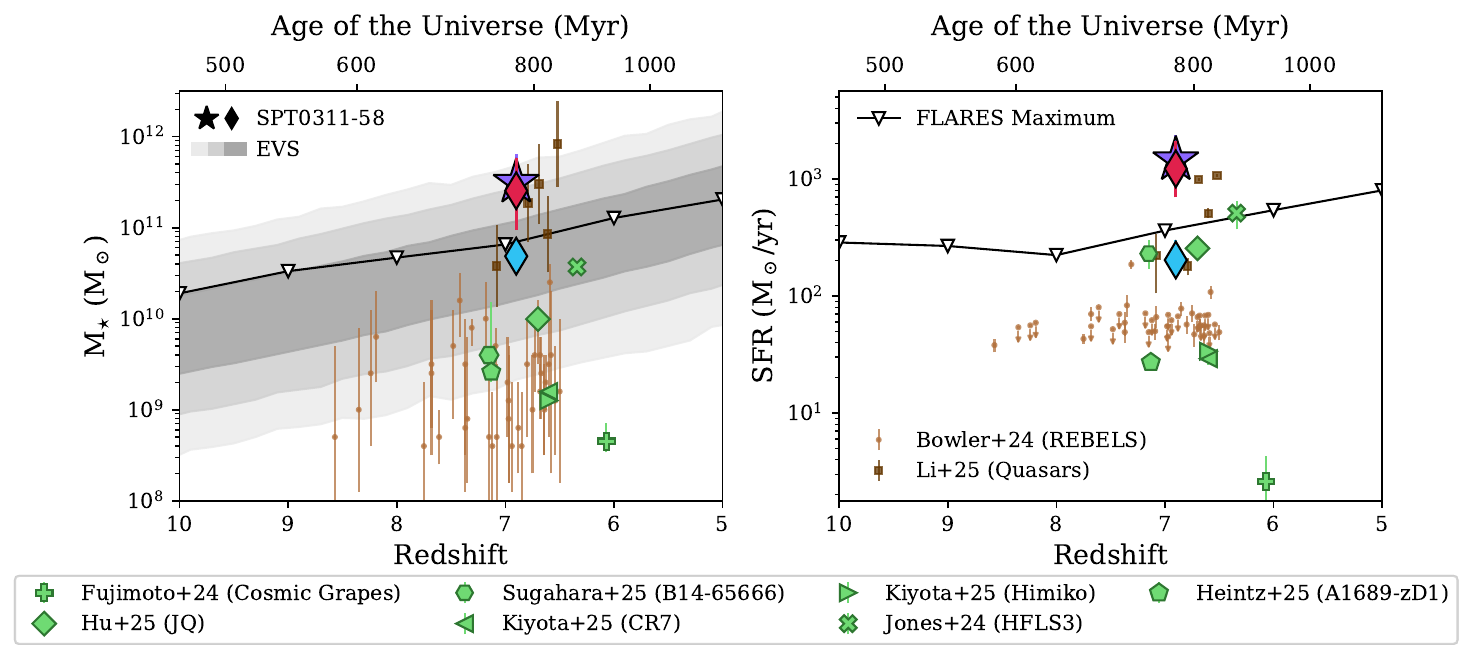}
    \caption{
    We compare the stellar mass (left) and SFR (right) of SPT0311$-$58 to other high-redshift ``extreme'' galaxies, including $z>6$ quasars (squares), bright LBGs from the REBELS survey (circles), and additional $z=6-7$ merging and/or clumpy systems (green symbols; \citealt{jones_ga-nifs_2024}, \citealt{fujimoto_primordial_2024}, \citealt{hu_extended_2025}, \citealt{sugahara_rioja_2025}, \citealt{kiyota_comprehensive_2025}, \citealt{heintz_inefficient_2025}); see text for details on these comparison objects. The whole of SPT0311$-$58 is represented as the purple star, SPT0311$-$58E as the blue diamond, and SPT0311$-$58W as the red diamond. White points show the most massive (left) and most highly star-forming (right) object from the FLARES cosmological simulation \citep{lovell_first_2021}, while the shaded region in the left panel illustrates the 68, 95, and 99.7\% confidence intervals for the most massive galaxy expected over a survey area of 25\,deg$^2$ \citep{marrone_galaxy_2018,lovell_extreme_2023}.
    }
    \label{fig:sims}
\end{figure*}

SPT0311$-$58 was deemed an extreme system by \citetalias{marrone_galaxy_2018} based on its exceptionally high SFR and inferred stellar mass. Here, we will compare SPT0311$-$58 with other systems that have historically been identified as extreme, both observationally and from simulations.

Our theoretical comparison includes two approaches. One is FLARES, a suite of cosmological zoom-in simulations of overdensities in the EoR \citep{lovell_first_2021, vijayan_first_2021}. The FLARES simulation is a well-suited point of comparison for SPT0311$-$58 as it was designed to target extreme systems within simulations. The other theoretical approach is based on a simple stellar mass-to-halo mass conversion and uses extreme value statistics \citep[EVS;][]{Gumbel_1958, Kotz_Nadarajah_2000}. We have adopted the EVS workflow as outlined in \citet{lovell_extreme_2023}. EVS assumes a DM halo distribution for the initial halo sample \citep{behroozi_average_2013}, followed by an assumed stellar mass distribution and baryonic fraction ($f_b$) to convert from halo to stellar masses. We assumed a baryon fraction of $f_b=0.16$ and a log-normal stellar mass distribution. We also adopted an effective survey area of 25\,deg$^2$, since SPT0311$-$58 was originally found in the 2500\,deg$^2$ SPT-SZ survey, but only $\sim$\,1\,$\%$ of foreground elliptical galaxies are expected to lens and magnify a source by a factor of two or more \citep{hezaveh_effects_2011, weis_alma_2013}. We utilized the \textsc{Python} module \texttt{EVSTATS} \citep{lovell_extreme_2023} along with the publicly available EVS probability density grid for this analysis\footnote{https://github.com/christopherlovell/evstats}.

Our observational comparison focuses on the following categories of extreme systems: high-redshift massive galaxies, $z>6$ quasars, and $z\sim6$--$7$ merging and/or clumpy systems. For the massive galaxies, we are comparing to the LBGs as part of the Reionization Era Bright Emission Line Survey (REBELS) program presented in \citet{bowler_alma_2024}, where they reported stellar masses and SFRs from $z$\,$=$\,6.5--8. We included the following five $z>6$ quasars from \cite{li_dichotomy_2025}, which used \jwst to measure the stellar masses, that have ALMA-informed SFRs: J0109-3047 \citep{meyer_alma_2023}, J0305-3150 \citep{meyer_pushing_2025}, J0224-4711 \citep{wang_spatially_2024}, J2002-3013 \citep{meyer_pushing_2025}, and J1120+0641 \citep{venemans_compact_2017}. SPT0311$-$58 is a unique system comprised of two massive, clumpy galaxies that are potentially merging. Only a handful of comparable $z\sim6$--$7$ systems have been discovered to date, including: 

\begin{itemize}
    \item[\smolbullet] HFLS3 \citep{riechers_dust-obscured_2013, jones_ga-nifs_2024}, another massive DSFG.
    \item[\smolbullet] \jwst's Quintet \citep[JQ;][]{hu_extended_2025}, a confirmed multi-merger system.
    \item[\smolbullet] Cosmic Grapes \citep{fujimoto_primordial_2024}, a single fragmented galaxy containing several star-forming clumps.
    \item[\smolbullet] B14-65666 \citep{sugahara_rioja_2025}, a dusty major merger.
    \item[\smolbullet] CR7 and Himiko \citep{kiyota_comprehensive_2025}, two additional clumpy systems.
    \item[\smolbullet] A1689-zD1 \citep{heintz_inefficient_2025}, a supposed ``normal'' DSFG from the EoR.
\end{itemize}

We show both the theoretical and observational comparisons to SPT0311$-$58 in Figure \ref{fig:sims}, where we plot stellar mass and SFR against redshift. SPT0311$-$58 as a whole is the most massive and most star-forming of the additional $z\sim6$--$7$ systems considered here. SPT0311$-$58E is forming stars at a similar rate to B14-65666 and JQ, while HFLS3 falls just short of SPT0311$-$58W.

As seen in the left panel of Figure \ref{fig:sims}, SPT0311$-$58 is much more massive than the LBGs in the REBELS sample, but is rather comparable in mass with the high-$z$ quasars. SPT0311$-$58 as a whole is more massive than the most massive galaxy in the FLARES simulation at $z\sim7$. Based on our EVS prediction, SPT0311$-$58 is consistent with the most massive galaxy expected from a 25\,deg$^2$ survey; however it lies in the upper $3\sigma$ of this distribution.

\citet{carnall_jwst_2024} performed a similar EVS analysis on a few massive quiescent galaxies from the Early eXtragalactic Continuum and Emission Line Science (EXCELS) survey. Using 160 arcminutes$^2$ as their effective survey area, they found that three galaxies (EXCELS-117560, EXCELS-109760, and UDS-7329) exceeded their EVS predicted mass at $z\gtrsim8$. When compared to the fiducial EVS model presented in \citet{carnall_jwst_2024}, the SFHs for these three EXCELS galaxies suggest a progenitor mass that lies in or beyond the upper $3\sigma$ regime. Historically, SFHs assume all stars formed within a single galaxy, but recent studies have found this assumption to overestimate stellar masses during early phases of galaxy formation. Formation histories consisting of multiple merger events, however, would alleviate this tension as stellar mass is built up from separate galaxies \citep{cochrane_hierarchical_2025}. This narrative, along with our EVS analysis, agrees with our picture of SPT0311$-$58, suggesting that SPT0311$-$58 is an example of a massive progenitor for these massive EXCELS galaxies.

\begin{figure*}[ht!]
    \centering
    \includegraphics[width=\linewidth]{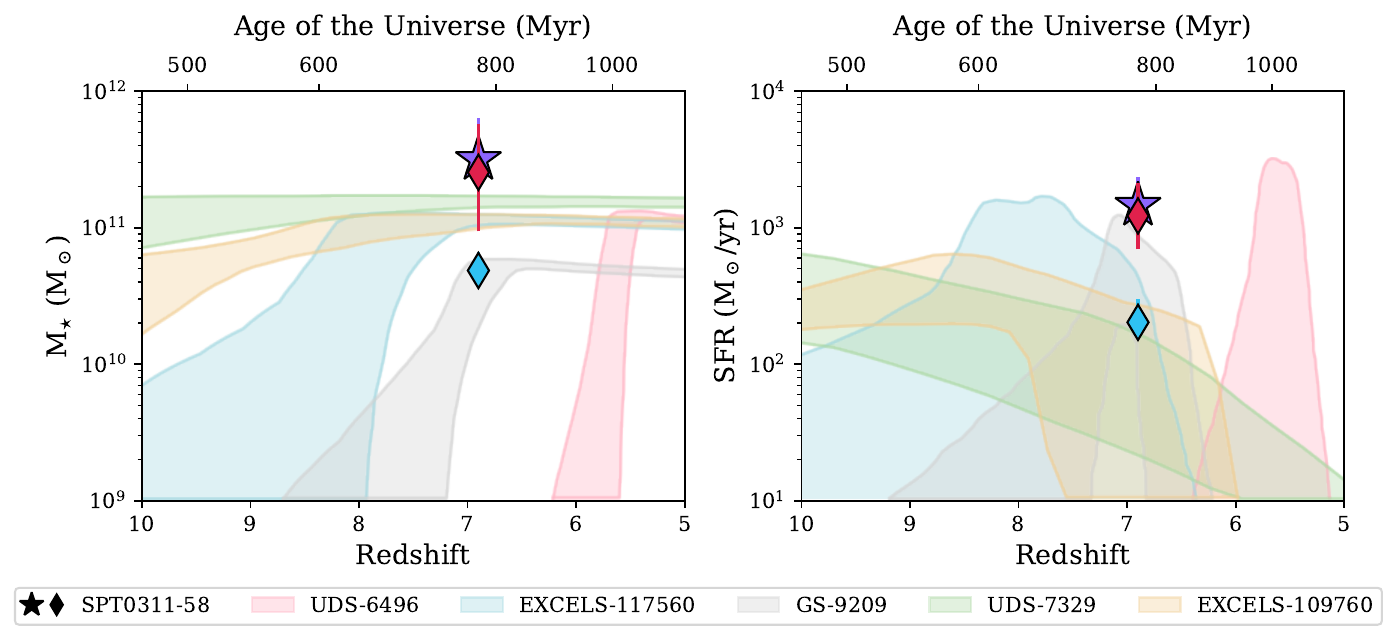}
    \caption{Comparing the stellar mass and SFR of SPT0311$-$58 as a whole (purple star), SPT0311$-$58E (blue diamond), and SPT0311$-$58W (red diamond) to the SFHs of the five massive $z\sim4$ quiescent galaxies considered here from the EXCELS survey \citep{carnall_jwst_2024, carnall_massive_2023, glazebrook_massive_2024}. As a massive star-forming galaxy, SPT0311$-$58 appears to be consistent with the SFHs of the highest stellar mass EXCELS galaxies at this epoch. The 1$\sigma$ confidence intervals for the SFHs were adapted from Figure 4 in \citet{carnall_jwst_2024}.}
    \label{fig:excels}
\end{figure*}

In the right panel of Figure \ref{fig:sims}, which shows SFR versus redshift, SPT0311$-$58 again falls above the FLARES simulation and well above the REBELS galaxies. Interestingly, it is again on par with QSOs in terms of SFR. Following the galaxy evolution model presented in \citet{hopkins_cosmological_2008}, DSFGs and QSOs are distinct but related evolutionary phases. SPT0311$-$58-like systems and high-$z$ QSOs may therefore represent distinct classes of massive galaxies that inhabit similar extremely massive DM halos. We currently have no evidence for an AGN in SPT0311$-$58, but we also cannot rule out the possibility of one or more AGN are highly dust-obscured or in an ``off'' phase of the duty cycle. Therefore, any potential evolutionary connection between SPT0311$-$58 and high-z QSOs is necessarily very speculative.

Aside from the similarity to high-$z$ QSOs, SPT0311$-$58 appears to be an extreme system for its redshift when compared to similar $z\sim6$--$7$ systems and high-$z$ LBGs, exceptionally so with respect to its SFR. While the starburst interpretation is debated given our earlier discussions about the SFH and SFMS, the incredibly high SFR may be driven by interactions between the individual clumps, particularly the unresolved dusty clumps in SPT0311$-$58W. This result underscores both the rarity of SPT0311$-$58 and the serendipity of its discovery.

\subsection{Relation to z$\sim$4 Quiescent Galaxies}

Here, we compare SPT0311$-$58 to the formation histories of massive $z\sim4$ quiescent galaxies to assess its viability as a progenitor to such systems. In Figure \ref{fig:excels}, we plot the mass and SFR of SPT0311$-$58 alongside the {\sc bagpipes}-derived SFHs of the five ultra-massive quiescent galaxies from the EXCELS survey \citep{carnall_jwst_2024}. We briefly mention the impact of abundance and survey areas, but we do not go into detailed consideration here.

At $z=6.9$, Figure \ref{fig:excels} shows that SPT0311$-$58, as a whole, is consistent at $1\sigma$ with the expected mass of the three most massive galaxies from the EXCELS sample (EXCELS-117560, -109760, and UDS-7329), but lies well above these galaxies in ongoing star-formation. Such an intense burst typically happens on a rather short timescale of $<$\,100\,Myr \citep{toft_submillimeter_2014}, a feature that would be washed out by the EXCELS SFHs. SPT0311$-$58E, on the other hand, has an SFR consistent with that inferred from the SFH of GS-9209. SPT0311$-$58E and W could likely evolve into separate massive quiescent galaxies by $z\sim3$. However, if the central galaxies were to merge, the resulting galaxy would have a stellar mass of $\log(M_\star/$M$_\odot)\gtrsim11.5$, which is on par with the most massive quiescent galaxies seen today \citep{ma_massive_2014}. 

SPT0311$-$58 is a rare source since it was the only source of its kind found within an effective survey area of 25\,deg$^2$. Not counting GS-9209, the four EXCELS galaxies were found in a survey area $\sim0.1\%$ of the area SPT0311$-$58 was found in. Given this abundance, the scarcity of systems like SPT0311$-$58 implies that, while they are the leading progenitors to date, systems like SPT0311$-$58 cannot be the sole progenitors of massive $z\gtrsim4$ quiescent galaxies.

Building off of the work from \citetalias{marrone_galaxy_2018}, \citet{jarugula_molecular_2021} used non-local thermodynamic equilibrium radiative transfer models to measure the gas mass with high-resolution CO and \ci ALMA observations. They found the gas mass to be $(3.1 \pm 2.7)\times10^{10}$\,M$_\odot$ for SPT0311$-$58E and $(5.4\pm 3.4)\times10^{11}$\,M$_\odot$ for SPT0311$-$58W. Within uncertainty, these are the same as our stellar masses presented here, which is typical of SMGs \citep{bothwell_survey_2013, aravena_survey_2016}. Using our SFRs and the gas masses from \citet{jarugula_molecular_2021}, the gas depletion times for SPT0311$-$58E and W are $160^{+64}_{-110}$\,Myr and $450^{+50}_{-85}$\,Myr, respectively. Both SPT0311$-$58E and W can therefore quench by $z\sim5$, further showing that this system could plausibly evolve into the massive quiescent galaxies seen between $z=$3--4.

The galaxies from the EXCELS survey mentioned here have stellar masses of $\log(M_\star/$M$_\odot)<11.5$ by the time of observation and are therefore assumed to have significant rotation \citep{carnall_jwst_2024}. However, there is no kinematic analysis yet to confirm this is the case. Recently, a massive ($\log(M_\star/$M$_\odot)\approx 11.5$) quiescent galaxy at $z$\,$\sim$\,3.5 (XMM-2599) has been confirmed as a slow-rotator \citep{forrest_massive_2025}. This system exhibits peculiar asymmetric morphology rather than the typical uniformity of massive elliptical galaxies. Being an asymmetric slow-rotator suggests that the early formation history was dominated by major mergers \citep{forrest_massive_2025}.

Using spectroscopic observations with MOSFIRE on Keck, \citet{forrest_extremely_2020} first hypothesized that DSFGs could be viable progenitors to XMM-2599, with SPT0311$-$58 being one of them. The SFH of XMM-2599 predicts $1\sigma$ upper mass limits of $\log(M_\star/$M$_\odot) \lesssim 11$ and SFR\,$\approx$\,$1000\,$M$_\odot\,\text{yr}^{-1}$ at $z\sim7$. Our new measurements of SPT0311$-$58 are about a factor of two above XMM-2599 for mass, but consistent in SFR. Even though SPT0311$-$58 is slightly more massive than the mass of XMM-2599 at $z=7$, SPT0311$-$58 overall aligns with the SFH of XMM-2599 well.

\section{Conclusions} \label{sec:conclusions}

We have presented new \jwst NIRCam and MIRI imaging of SPT0311$-$58, a clumpy starburst system at $z$\,$=$\,6.9. Supplementing these with data from \hst and \jwst NIRSpec, along with ALMA dust measurements, we have constrained stellar masses and SFRs on clumpy and global scales.

\begin{itemize}

\item[\smolbullet] By combining dust and UV information, we have shown that the clumps in SPT0311$-$58E vary greatly in dust obscuration. Nearly the entire stellar component of SPT0311$-$58W is completely dust-obscured; W8 is the only clump to have a UV counterpart (C3). 

\item[\smolbullet] Nearly all individual regions considered here behave as typical $z=6.9$ star-forming galaxies, but the incredibly high SFR from SPT0311$-$58W pushes the overall system well above the SFMS. Our stellar masses and SFRs agree with previously reported results using SED fitting and/or dust emission (\citetalias{marrone_galaxy_2018}, \citetalias{alvarez-marquez_mirijwst_2023}). We confirm the SFRs presented in \citetalias{arribas_ga-nifs_2024} derived using \halpha are significantly lower for the dustier galaxy (SPT0311$-$58W). 

\item[\smolbullet] SPT0311$-$58 lies about an order of magnitude in mass and SFR above other high-redshift massive galaxies from the REBELS survey. We do find that SPT0311$-$58 is comparable to $z>6$ quasars in stellar mass and SFR, even though it lacks any evidence for an AGN. SPT0311$-$58 is more massive and star-forming than the respective maximums from the FLARES simulations. We also find that SPT0311$-$58 as a whole lies within the $3\sigma$ upper limit of our EVS predictions for a survey area of 25\,deg$^2$.

\item[\smolbullet] We compared SPT0311$-$58 with the SFHs of a few massive quiescent galaxies from the EXCELS survey. When considering the collective SFHs, we find that SPT0311$-$58 is consistent in stellar mass with the most massive galaxies considered here, but falls above the respective SFRs. The SFHs likely blur out intense SFR episodes, like the one seen with SPT0311$-$58, so this difference is not cause for concern. Therefore, SPT0311$-$58 is an example of a viable progenitor to such systems. 

\end{itemize}
 
\FloatBarrier 

\begin{acknowledgments}
    BW thanks the anonymous referee for their constructive feedback for improving this manuscript. The authors thank Ryan Endsley, Christopher Lovell, and Daniel Stark for helpful discussions about the proposal for program \#1791 and subsequent data reduction.
    
    This work is based on observations made with the NASA/ESA/CSA \jwst. The data were obtained from the Mikulski Archive for Space Telescopes at the Space Telescope Science Institute, which is operated by the Association of Universities for Research in Astronomy, Inc., under NASA contract NAS 5-03127 for JWST.  BW, JEB, and JSS acknowledge funding support associated with program \#1791.
    
    The SPT is supported by the NSF through grant OPP-1852617. This work was partially supported by the Center for AstroPhysical Surveys (CAPS) at the National Center for Supercomputing Applications (NCSA), University of Illinois Urbana-Champaign. MA is supported by FONDECYT grant number 1252054, and gratefully acknowledges support from ANID Basal Project FB210003 and ANID MILENIO NCN2024\_112.

    This paper makes use of the following ALMA data: ADS/JAO.ALMA\#2016.1.01293.S, ADS/JAO.ALMA\#2017.1.01423.S. ALMA is a partnership of ESO (representing its member states), NSF (USA) and NINS (Japan), together with NRC (Canada), MOST and ASIAA (Taiwan), and KASI (Republic of Korea), in cooperation with the Republic of Chile. The Joint ALMA Observatory is operated by ESO, AUI/NRAO and NAOJ.  This research has made use of NASA’s Astrophysics Data System and Astrophysics Data System Bibliographic Services.
\end{acknowledgments}

\facilities{\hst(ACS and WFC3), \jwst(NIRSpec, NIRCam, and MIRI), ALMA}

\software{\texttt{astropy} \citep{2013A&A...558A..33A,2018AJ....156..123A,2022ApJ...935..167A},
          \textsc{cigale} \citep{noll_analysis_2009,boquien_cigale_2019}, 
          \texttt{galight} \citep{ding_galaxy_2021},
          \texttt{lenstronomy} \citep{birrer2018lenstronomymultipurposegravitationallens,2021JOSS....6.3283B},
          \textsc{magphys} \citep{da_cunha_simple_2008, da_cunha_alma_2015},
          \texttt{matplotlib} \citep{hunter_matplotlib_2007},
          \texttt{numpy} \citep{harris_array_2020},
          \textsc{Python},
          \texttt{webbpsf} \citep{perrin_updated_2014}
          }

\bibliography{bibitems0311}{}
\bibliographystyle{aasjournal}

\appendix

\section{Profile Fitting Photometry}

We include the results of our profile-fitting photometry (see Section \ref{subsec:profilephot}) in Table \ref{tab:phot}.

\noindent
\begin{centering}
\begin{table*}[ht]
\setlength{\tabcolsep}{12pt}
\tiny
\begin{tabular}{llllllc}
\hline\hline
Region & \multicolumn{5}{c}{Filters} & Instr. \\
\hline
& F606W  & F775W  & F814W  & F125W  & F160W & \multirow{15}{*}{\rotatebox{-90}{\hspace{5mm} \hst ACS \& WFC3}} \\
\cline{2-6}
L1 & $0.023 \pm 0.014$ & $ < 0.0099$ & $ < 0.0039$ & $ < 0.0079$ & $0.018 \pm 0.0064$ \\[0.8ex]
L2 & $ < 0.069$ & $ < 0.046$ & $ < 0.017$ & $ < 0.033$ & $ < 0.025$ \\[0.8ex]
L3 & $ < 0.032$ & $ < 0.022$ & $ < 0.0094$ & $ < 0.026$ & $ < 0.021$ \\[0.8ex]
L4 & $ < 0.045$ & $ < 0.031$ & $ < 0.012$ & $ < 0.03$ & $ < 0.027$ \\[0.8ex]
L5 & $ < 0.038$ & $ < 0.026$ & $ < 0.01$ & $ < 0.02$ & $ < 0.016$ \\[0.8ex]
L6 & $ < 0.032$ & $ < 0.021$ & $ < 0.0083$ & $0.031 \pm 0.017$ & $0.017 \pm 0.013$ \\[0.8ex]
C2 & $ < 0.1$ & $ < 0.072$ & $ < 0.029$ & $ < 0.085$ & $ < 0.062$ \\[0.8ex]
ENc & $0.017 \pm 0.0093$ & $ < 0.0064$ & $ < 0.0025$ & $0.0055 \pm 0.0053$ & $0.018 \pm 0.0039$ \\[0.8ex]
E1+C1 & $ < 0.11$ & $ < 0.075$ & $ < 0.029$ & $ < 0.066$ & $ < 0.06$ \\[0.8ex]
E2+ENb & $ < 0.011$ & $0.03 \pm 0.0077$ & $ < 0.0031$ & $0.018 \pm 0.008$ & $0.033 \pm 0.0062$ \\[0.8ex]
E3+ENa & $ < 0.041$ & $ < 0.029$ & $ < 0.012$ & $ < 0.029$ & $ < 0.024$ \\[0.8ex]
E4+ES & $ < 0.12$ & $ < 0.085$ & $ < 0.034$ & $ < 0.1$ & $ < 0.08$ \\[0.8ex]
W1-7 & $ < 0.16$ & $ < 0.11$ & $ < 0.047$ & $ < 0.11$ & $ < 0.092$ \\[0.8ex]
W8+C3 & $ < 0.064$ & $ < 0.044$ & $ < 0.018$ & $ < 0.04$ & $ < 0.031$ \\[0.8ex]
\hline
& F200W  & F300M  & F335M  & F444W & & \multirow{15}{*}{\rotatebox{-90}{\hspace{5mm} \jwst NIRCam}}\\
\cline{2-5}
L1 & $0.026 \pm 0.0048$ & $0.064 \pm 0.0058$ & $0.046 \pm 0.006$ & $0.1 \pm 0.0076$ \\[0.8ex]
L2 & $ < 0.0076$ & $ < 0.0091$ & $ < 0.0096$ & $0.019 \pm 0.015$ \\[0.8ex]
L3 & $ < 0.0092$ & $ < 0.013$ & $ < 0.013$ & $ < 0.018$ \\[0.8ex]
L4 & $ < 0.014$ & $ < 0.017$ & $ < 0.017$ & $ < 0.023$ \\[0.8ex]
L5 & $ < 0.012$ & $0.035 \pm 0.013$ & $0.025 \pm 0.014$ & $0.07 \pm 0.019$ \\[0.8ex]
L6 & $0.023 \pm 0.0097$ & $0.033 \pm 0.011$ & $0.031 \pm 0.012$ & $0.041 \pm 0.016$ \\[0.8ex]
C2 & $0.083 \pm 0.029$ & $0.16 \pm 0.035$ & $0.19 \pm 0.035$ & $0.24 \pm 0.052$ \\[0.8ex]
ENc & $0.023 \pm 0.0035$ & $0.068 \pm 0.0043$ & $0.077 \pm 0.0044$ & $0.1 \pm 0.0052$ \\[0.8ex]
E1+C1 & $0.068 \pm 0.03$ & $0.13 \pm 0.034$ & $0.12 \pm 0.035$ & $0.2 \pm 0.051$ \\[0.8ex]
E2+ENb & $0.058 \pm 0.0034$ & $0.16 \pm 0.0041$ & $0.14 \pm 0.0042$ & $0.23 \pm 0.0055$ \\[0.8ex]
E3+ENa & $0.077 \pm 0.013$ & $0.14 \pm 0.017$ & $0.12 \pm 0.018$ & $0.2 \pm 0.022$ \\[0.8ex]
E4+ES & $0.14 \pm 0.033$ & $0.27 \pm 0.041$ & $0.29 \pm 0.041$ & $0.39 \pm 0.06$ \\[0.8ex]
W1-7 & $ < 0.047$ & $ < 0.061$ & $ < 0.062$ & $ < 0.085$ \\[0.8ex]
W8+C3 & $0.06 \pm 0.024$ & $0.15 \pm 0.035$ & $0.16 \pm 0.035$ & $0.24 \pm 0.041$ \\[0.8ex]
\hline
& F560W  & F770W  & F1000W  & F1280W  & F1800W & \multirow{15}{*}{\rotatebox{-90}{\hspace{5mm} \jwst MIRI}}\\
\cline{2-6}
L1 & $0.086 \pm 0.041$ & $0.15 \pm 0.057$ & $ < 0.078$ & $ < 0.19$ & $ < 0.54$ \\[0.8ex]
L2 & $ < 0.18$ & $ < 0.26$ & $ < 0.34$ & $ < 0.87$ & $ < 2.2$ \\[0.8ex]
L3 & $ < 0.12$ & $ < 0.19$ & $ < 0.32$ & $ < 1.1$ & $ < 5.9$ \\[0.8ex]
L4 & $ < 0.14$ & $ < 0.2$ & $ < 0.29$ & $ < 0.85$ & $ < 4.0$ \\[0.8ex]
L5 & $ < 0.11$ & $ < 0.15$ & $ < 0.2$ & $ < 0.49$ & $ < 1.4$ \\[0.8ex]
L6 & $ < 0.092$ & $ < 0.14$ & $ < 0.19$ & $ < 0.51$ & $ < 1.5$ \\[0.8ex]
C2 & $ < 0.34$ & $ < 0.56$ & $ < 0.78$ & $ < 2.4$ & $ < 11$ \\[0.8ex]
ENc & $0.18 \pm 0.026$ & $0.2 \pm 0.037$ & $0.33 \pm 0.049$ & $0.41 \pm 0.13$ & $3.4 \pm 0.46$ \\[0.8ex]
E1+C1 & $ < 0.34$ & $ < 0.49$ & $ < 0.68$ & $ < 1.8$ & $ < 4.9$ \\[0.8ex]
E2+ENb & $0.23 \pm 0.034$ & $0.23 \pm 0.053$ & $0.34 \pm 0.08$ & $0.69 \pm 0.23$ & $ < 0.8$ \\[0.8ex]
E3+ENa & $0.26 \pm 0.15$ & $0.37 \pm 0.24$ & $0.87 \pm 0.39$ & $ < 1.3$ & $ < 7.1$ \\[0.8ex]
E4+ES & $ < 0.4$ & $ < 0.62$ & $ < 0.93$ & $ < 2.9$ & $ < 13$ \\[0.8ex]
W1-7 & $0.61 \pm 0.55$ & $2.7 \pm 0.87$ & $1.6 \pm 1.4$ & $8.7 \pm 4.5$ & $ < 25$ \\[0.8ex]
W8+C3 & $0.37 \pm 0.21$ & $ < 0.32$ & $ < 0.45$ & $ < 1.3$ & $ < 4.8$ \\[0.8ex]
\hline\hline
\end{tabular}
\caption{Measured photometry (in $\mu$Jy) for the 14 filters used in our multi-band fitting routine using \texttt{galight}. We report the $1\sigma$ confidence intervals for the uncertainties and limits. The table has been divided by telescope and/or instrument.}
\label{tab:phot}
\end{table*}
\end{centering}

\end{document}